\documentclass[journal, onecolumn]{IEEEtran}
\usepackage[utf8x]{inputenc} 
\usepackage{amsmath,graphicx}
\usepackage{amssymb}
\usepackage{algorithm}
\usepackage[noend]{algpseudocode}
\usepackage{enumitem}
\usepackage[list=true]{subcaption}
\usepackage{upgreek}
\graphicspath{{RESULTS/}} %Setting the graphicspath
\usepackage{bm}
\usepackage{booktabs} 
\def\X{{\mathbf X}}

\def\C{{\mathbb C}}
\def\R{{\mathbb R}}

\def\E{{\mathbb E}}

\def\SNR{\mathrm{SNR}}
\def\RMSE{\mathrm{RMSE}}

\DeclareMathOperator*{\argmin}{arg\,min}
\renewcommand{\Re}{\operatorname{Re}}
\renewcommand{\Im}{\operatorname{Im}}

\newcommand{\ju}{\mathrm j}
\usepackage{hyperref}

\newcommand{\RANK}{\operatorname{rank}}

\newcommand{\ang}{\bm\uptheta}

\DeclareMathOperator{\diag}{diag}

\newcommand{\Saug}{\mathbf{\bar{S}}}
\newcommand{\Eaug}{\mathbf{\bar{E}}}
\usepackage{hyperref}

\begin{document}

\title{Deep Networks for Direction-of-Arrival Estimation\\in Low SNR}

\author{Georgios~K.~Papageorgiou,~\IEEEmembership{Member,~IEEE,}, Mathini~Sellathurai,~\IEEEmembership{Senior Member,~IEEE,}  
		and~Yonina~C.~Eldar,~\IEEEmembership{Fellow,~IEEE}}

%\markboth{Journal of \LaTeX\ Class Files,~Vol.~14, No.~8, August~2020}%

\maketitle

\begin{abstract}
In this work, we consider direction-of-arrival (DoA) estimation in the presence of extreme noise using Deep Learning (DL). In particular, we introduce a Convolutional Neural Network (CNN) that is trained from mutli-channel data of the true array manifold matrix and is able to predict angular directions using the sample covariance estimate. We model the problem as a multi-label classification task and train a CNN in the low-SNR regime to predict DoAs across all SNRs. The proposed architecture demonstrates enhanced robustness in the presence of noise, and resilience to a small number of snapshots. Moreover, it is able to resolve angles within the grid resolution. Experimental results demonstrate significant performance gains in the low-SNR regime compared to state-of-the-art methods and without the requirement of any parameter tuning. We relax the assumption that the number of sources is known a priori and present a training method, where the CNN learns to infer the number of sources jointly with the DoAs. Simulation results demonstrate that the proposed CNN can accurately estimate off-grid angles in low SNR, while at the same time the number of sources is successfully inferred for a sufficient number of snapshots. Our robust solution can be applied in several fields, ranging from wireless array sensors to acoustic microphones or sonars. 
\end{abstract}

\begin{IEEEkeywords}
direction-of-arrival DoA estimation, array signal processing, multilabel classification, deep learning, convolution neural network CNN
\end{IEEEkeywords}

\IEEEpeerreviewmaketitle

\section{Introduction}
\label{sec:intro}

\IEEEPARstart{D}{irection}-of-arrival (DoA) estimation has been at the forefront of research activity for many decades, due to the plethora of applications ranging from radar and wireless communications to sonar and acoustics \cite{zoubir2013academic}, with localization being one of the most significant ones. Estimation of the angular directions is possible with the use of multiple sensors in a specified geometric configuration, e.g., linear, rectangular and circular. Efficient use of the observations from multiple sensors enables the DoA estimation of several sources, depending on the number of array sensors. There are two major problem categories in DoA estimation: the overdetermined case, where  the number of sources is less than the number of array sensors, and the underdetermined case, where the number of sources is equal or greater than the number of sensors \cite{7055290, 5456168}. In this work we focus on the first category (extension to the underdetermined case is straightforward with minor modifications)\footnote{In the underdetermined case non-uniform array configurations are typically utilized.}. 

One of the first methods (also popular to this day) that was introduced for DoA estimation is MUltiple SIgnal Classification (MUSIC) \cite{SchmidtMUSIC}, with several other variants following soon after. The MUSIC estimator belongs to the class of subspace-based techniques, since it attempts to separate the signal and noise sub-spaces; the angle estimation follows from the so-called MUSIC pseudo-spectra over a specified grid, where the corresponding peaks of the pseudo-spectra are selected. Therefore, it is considered a grid-based DoA estimation technique. Estimation of signal parameters via rotational invariance techniques (ESPRIT) \cite{32276} is another successful example, but requires two physical arrays for DoA estimation. A significant step towards improvement in DoA estimation was with the development of another subspace-based variant, the Root-MUltiple SIgnal Classification (R-MUSIC) algorithm, which estimates the angular directions from the solutions of higher-order polynomials \cite{1172124}. The advantage of this method is that no grid is required, therefore, it leads to more accurate DoA estimates in both the high and the low signal-to-noise-ratio (SNR) regime for a sufficient number of snapshots. The aforementioned methods, are covariance-based techniques that require a sufficient number of data snapshots to accurately estimate the DoAs, particularly in the low-SNRs. In addition, they are sensitive to the angular separation of the sources. If the angles are not sufficiently separated, their performance degrades significantly. Furthermore, they typically assume that the number of sources is known, which is not the case in many practical applications.

During the past decade, another approach has emerged from the sparse representation and Compressed Sensing (CS) methodology \cite{1614066, eldar2012compressed}. These methods exploit the sparse characteristic of the signal sources in the spatial domain (angles), thus, sparse signal recovery approaches can be applied. Furthermore, under certain conditions, such as the restricted-isometry-property (RIP), the sparse signal can be stably reconstructed with the reconstruction error being proportional to the noise level \cite{candes2008restricted}. In general, they are separated into three main categories: a) on-grid, b) off-grid and c) grid-less methods \cite{yang2018sparse}, \cite{4014378}. Grid-less methods achieve better performance at the expense of very high computational complexity, which can be prohibitive in many practical applications \cite{6857424}. Off-grid and on-grid methods offer a more balanced solution with lower computational complexity at the expense of negligible loss in performance. However, the DoA estimates are obtained after solving computationally demanding optimization tasks, e.g., in the Least Absolute Shrinkage and Selection Operator (LASSO) or Basis-pursuit Denoising (BPDN) forms (which involve the minimization of the mixed $\ell_{2,1}$ norm). One of the major disadvantages that all these methods share in common is that the tuning of one or more parameters (which depends on the number of snapshots, the SNR or both) is required to guarantee good performance; moreover, the DoA estimates are often extremely sensitive to the tuning of these parameters.  

A very recent approach to DoA estimation is via the use of Deep Learning (DL) \cite{lecun2015deep,goodfellow2016deep}. A deep neural network (DNN) with fully connected (FC) layers was employed in \cite{8555814} for DoA classification of two targets using the signal covariance matrix. However, the reported results indicate poor DoA estimation results in the high SNR. The authors in \cite{8400482} proposed another DNN for channel estimation in massive MIMO systems; nevertheless, their work is in the direction of a) DoA tracking to learn the communication channels and b) high-SNR regime operation. A multilayer autoencoder with a series of parallel multilayer classifiers, i.e., a multi-layer perceptron (MLP), was employed in \cite{8485631} with focus on the robustness to array imperfections. A deep Convolutional Neural Network (CNN) that was also trained in low SNRs was proposed in \cite{8854868}. However, the method did not demonstrate significant performance improvement in terms of DoA estimation, due to the adoption of 1-dimensional (1D) filters (convolutions). To the best of our knowledge \emph{no specific DL-based technique was designed for DoA estimation in the low-SNR regime}. Another disadvantage of previous methods such as \cite{8485631, 8854868} is that they are trained for a specific number of snapshots. This in turn, leads to a) significant deviations for different number of snapshots and b) to the necessity to re-train the network.

The scope of this work is to fill in the gap in the literature of DoA estimation in the low-SNR regime with the use of DL. Sample covariance matrix estimates in the low SNRs are characterized by large deviations from the true manifold matrix. Therefore, DoA estimation becomes really challenging and the majority of the methods fail to demonstrate the desired robustness. In this work, we contribute towards this direction by: a) using multi-channel data and b) exploiting 2-dimensional (2D) convolutional layers, which are well known for their excellent feature extraction properties. Moreover, DL-based methods have several advantages over optimization-based ones. First, after training the neural network no optimization is required and the solution is the result of simple operations (multiplications and additions). Additionally, they do not require any specific tuning of parameters, in contrast to optimization-based techniques. Last but not least, neural networks demonstrate robustness in terms of DoA estimation accuracy, e.g., using fewer snapshots, performing well in the low-SNR regime. In order to derive such automated solutions we need to identify a suitable network architecture and train it efficiently. For example, networks can be trained to either ``learn'' the (spatial) spectrum or the DoAs directly; they can be trained to make predictions at certain SNRs or across a range of SNRs. 

Our contributions are summarized as follows: 
\begin{itemize}
\item We introduce a deep CNN trained on multi-channel data, which are explicitly formed from the complex-valued data of the true covariance matrix. The proposed CNN employs 2-dimensional (2D) convolutional layers and is trained to directly predict the angular directions of multiple sources using the sample covariance estimate. The use of multi-channel data along with the adoption of 2D convolutional layers enables the extraction of features from the input data leading to a robust DoA estimation, especially in the low SNRs. A discretization approach is adopted for the desirable angular (spatial) region and the machine learning task is modeled as a multi-label classification one.
\item We present efficient methods for the training of the proposed network. In particular, we train the CNN across a range of low-SNRs and demonstrate that it can successfully predict DoAs in the high-SNR regime as well.
\item The assumption that the number of transmitting sources is known a priori is relaxed. To this end, we introduce a training method for a varying number of sources. Subsequently, \emph{the proposed CNN is able to infer an unknown number of DoAs leading to the simultaneous prediction of the number of sources}. 
\item The performance of the proposed solution is evaluated over an extensive set of simulated experiments, where it is compared against state-of-the-art methods in various experimental set-ups with off-grid angles. Additionally, comparison to the Cram\'{e}r-Rao lower bound (CRLB) is provided as benchmark. The results indicate that the proposed CNN: a) outperforms its competitors in DoA estimation in the low-SNR regime; b) is resilient in estimation even for a small collection of snapshots and regardless of the angular separation of the sources; c) demonstrates enhanced robustness in case of SNR mismatches, and
d) is able to infer both the number of sources and DoAs with very small errors and high confidence level. 
\end{itemize}

The rest of the paper is organized as follows: in Section \ref{sec:signal_model}, we present the signal model. Section \ref{sec:preliminaries} provides an overview of DoA estimators, explicitly selected for estimation in the low-SNR regime. In Section \ref{sec:DoA_est_CNN}, we introduce the proposed CNN for DoA prediction, whereas in Section\ref{sec:training_approach} we discuss the adopted training approach. Section \ref{sec:results} presents the simulation results and in Section \ref{sec:conclusions}, we summarize and highlight our conclusions.

\textbf{Notation}: Throughout the paper the following notation is adopted: $\mathcal{X}$ denotes a set and $| \mathcal{X} |$ its cardinality; $\X$ is a matrix, $\mathbf{x}$ is a vector and $x$ is a scalar. The $i,j$-th element of matrix $\X$ is denoted $( \X )_{i,j}$ and the $i$-th entry of vector $\mathbf{x}$ as $x(i).$ The $i$-th example of the vector $\mathbf{x}$ is denoted as $\mathbf{x}(i)$. The imaginary unit is $\ju$ (so that $\ju^2=-1$). The conjugate transpose of a matrix is $(\cdot)^H$; its conjugate is $(\cdot)^*$ and its transpose is $(\cdot)^T.$ The $N\times N$ identity matrix is $\mathbf{I}_N$. The white circularly-symmetric Gaussian distribution with mean $\mathbf{m}$ and covariance $\mathbf{C}$ is denoted by $\mathcal{CN}(\mathbf{m},\mathbf{C}).$ The notation $\|\cdot \|_F$ is the Frobenius norm of a matrix and $\|\cdot \|_{2,1}$ is the mixed $\ell_{2,1}$ norm. The convolution operator is denoted as $*$. The floor of a number $\alpha$ is written as $\lfloor \alpha \rfloor.$ Functions are denoted by lower case italics, e.g., $f(\cdot)$. The symbol $\E[\cdot]$ is the expectation operator; $\Re\{\cdot\}, \Im\{\cdot\}$ denote the real and the imaginary parts of a complex scalar / vector / matrix, respectively. Finally, the phase of the complex-valued variable $\alpha$ is denoted by $\angle \{\alpha\}$.

\section{Signal Model}
\label{sec:signal_model}

The standard model for a $N$-element sensor array in the narrow-band mode, with $K$ far-field sources present, is:
\begin{align}
\mathbf{y}(t) =& \sum_{k=1}^K \mathbf{a}(\theta_k)s_k(t)+\mathbf{e}(t)= \nonumber \\
=& \mathbf{A}(\ang)\mathbf{s}(t)+\mathbf{e}(t),\ t=1,\dots,T.
\label{eq:signal_model}
\end{align}
Here $\mathbf{A}(\ang)= [\mathbf{a}(\theta_1),\mathbf{a}(\theta_2),\dots, \mathbf{a}(\theta_K)]$ is the $N\times K$ array manifold matrix, $\ang = [\theta_1,\dots,\theta_K]^T$ is the vector of (unknown) source directions and $T$ is the total number of collected snapshots, $\mathbf{s}(t)=[s_1(t),\dots, s_K(t)]^T$ and $\mathbf{e}(t)$ denote the transmitted signal and additive noise vectors at sample index $t$, respectively. The model \eqref{eq:signal_model} is generic in the sense that it does not depend on the array geometry; however, in this work we will consider a uniform linear array (ULA) configuration for simplicity\footnote{The analysis and methodology also holds for any other array configuration, e.g.,  non-uniform linear or rectangular.}. Thus, the columns of the array manifold matrix can be expressed as 
\begin{equation}
\mathbf{a}(\theta_k)=[1, e^{ \ju \frac{2\pi d}{\lambda} \sin(\theta_k)}, \dots,e^{ \ju \frac{2\pi d}{\lambda} \sin(\theta_k)(N-1)}]^T,
\label{eq:array_mainfold_columns}
\end{equation}
where $d$ is the array interelement distance and $\lambda=c/f$ is the wavelength at carrier frequency $f$ with $c$ the speed of light. In this case, $\mathbf{A}(\ang)$ becomes a Vandermonde matrix. We make the following assumptions throughout the paper:
\begin{enumerate}[label=A\arabic*.]
\item The source DoAs are distinct. 
\item Each source signal follows the unconditional-model assumption (UMA) in \cite{Stioca1990}, which assumes that the transmitted signal is randomly generated (Gaussian signaling). Moreover, the sources are uncorrelated, leading to a diagonal source covariance matrix: $\mathbf{R}_s=\E[\mathbf{s}(t)\mathbf{s}^H(t)]=\diag(\sigma_1^2,\dots,\sigma_K^2)$.
\item The additive noise values are independent and identically distributed (i.i.d.) zero-mean white circularly-symmetric Gaussian, i.e., $\mathbf{e}(t)\sim \mathcal{CN}(\mathbf{0},\sigma_{e}^2 \mathbf{I}_N)$ and uncorrelated from the sources.
\item There is no temporal correlation between each snapshot. 
\end{enumerate}

We are interested in the estimation of the unknown DoAs $\ang$ from measurements $\mathbf{y}(1),\dots,\mathbf{y}(T).$ Under assumptions A1--A4, the received signal's covariance matrix is given by:
\begin{equation}
\mathbf{R}_y = \E[\mathbf{y}(t)\mathbf{y}^H(t)]=\mathbf{A(\ang)R}_s \mathbf{A}^H(\ang) + \sigma_{e}^2 \mathbf{I}_N.
\label{eq:theor_cov_matrix}
\end{equation}
The statistical richness of $\mathbf{R}_y$ in \eqref{eq:theor_cov_matrix} allows for the estimation of up to $K\leq N-1$ distinct DoAs. However, in practice, the matrix in \eqref{eq:theor_cov_matrix} is unknown and is replaced by its sample estimate
\begin{equation}
\widetilde{\mathbf{R}}_y = \frac{1}{T} \sum_{t=1}^T \mathbf{y}(t)\mathbf{y}^H(t),
\label{eq:sam_cov_matrix}
\end{equation}
which is an unbiased estimator of $\mathbf{R}_y$.

\section{DoA Estimation with\\ Multiple Measurement Vectors (MMV)}
\label{sec:preliminaries}
In this section we provide a brief summary of various MMV-based DoA estimators and discuss their pros and cons.

\subsection{Multiple Signal Classification (MUSIC)}
\label{ssec:MUSIC}

MUSIC is one the first algorithms developed and is among the most popular ones to date, belonging to the class of subspace-based methods. The estimator separates the signal and noise sub-spaces via the eigendecomposition of the covariance matrix $\mathbf{R}_y$ in \eqref{eq:theor_cov_matrix}, which is given by:
\begin{equation}
\mathbf{R}_y = \mathbf{Q \Lambda Q}^H,
\label{eq:spatially_smoothed_cov_mat_eigendecomp}
\end{equation}
where $\mathbf{\Lambda} = \diag (\lambda_1, \dots,\lambda_K, \lambda_{K+1}, \dots, \lambda_{N})$ is the diagonal matrix of eigenvalues in decreasing order\footnote{It is also straightforward to see that $\lambda_{K+1}=\dots=\lambda_{N}=\sigma_e^2$ and $\lambda_1>\dots>\lambda_K>\sigma_e^2$.} and $\mathbf{Q}=[\mathbf{Q}_s\ \mathbf{Q}_{e}]$ is an $N\times N$ eigenvector matrix whose first $K$ column vectors correspond to the signal subspace spanned by $\mathbf{Q}_s$, whereas the remaining $N-K$ columns correspond to the noise subspace spanned by $\mathbf{Q}_e \in \C^{N\times (N-K)}$. Due to the fact that the columns of a) $\mathbf{Q}_s$ and $\mathbf{Q}_e$ are orthogonal and b) $\mathbf{A}(\ang)$ and $\mathbf{Q}_s$ span the same space we have $\|\mathbf{A}^H(\ang)\mathbf{Q}_e\|_F=0.$ The MUSIC spectra can be calculated from: 
\begin{equation}
P_{\text{MUSIC}}(\phi_i) = \frac{1}{\mathbf{a}^H(\phi_i) \mathbf{Q}_{e}\mathbf{Q}_{e}^H \mathbf{a}(\phi_i)},\ \text{for }\ i=1,\dots,G_{\text{M}},
\label{eq:P_music}
\end{equation}
where $\mathbf{a}(\phi_i)$ is the response/steering vector of the array, $\phi_i \in \mathcal{G}$ is the discretized grid angle on $\mathcal{G}$ (according to the desired resolution) and $G_{\text{M}}$ is the number of grid points. Since the signal and noise subspaces are orthogonal to each other, the denominator in \eqref{eq:P_music} will become zero if one of the $\phi_i$'s is a true source DoA (or close to its true value). Hence, the MUSIC spectra will assume a peak, thus enabling the signal classification of $K$ sources with $K < N$. However, in practice, as we previously mentioned, the true covariance matrix is unknown; therefore, the aforementioned process is followed for $\widetilde{\mathbf{R}}_y$ instead of $\mathbf{R}_y,$ leading to an estimated noise subspace, $\widetilde{\mathbf{Q}}_e$. Particularly in the low-SNR regime, there is no clear distinction between the signal and noise eigenvalues often leading to the so-called subspace swap, which results in inaccurate DoA estimates. Another disadvantage of the method is the finite resolution, due to the grid points in $\mathcal{G}$ and the computational complexity of the repeatedly performed grid search.

\subsection{Root Multiple Signal Classification (R-MUSIC)}
\label{ssec:RMUSIC}

An improvement of the MUSIC algorithm is the Root Multiple Signal Classification (R-MUSIC) algorithm. The steering vector $\mathbf{a}(\theta)$ in \eqref{eq:array_mainfold_columns} can be expressed as a function of $\omega=e^{-j\frac{2\pi d}{\lambda}\sin(\theta)}$:
\begin{equation}
\mathbf{a}(\omega)=[1, \omega^{-1}, \omega^{-2},\dots,\omega^{-(M-1)}]^T.
\label{eq:rmusic_steering_vec}
\end{equation}
The standard R-MUSIC algorithm transforms the spectral search step involved in MUSIC into a simplified polynomial rooting of the following function 
\begin{equation}
f_{\text{R-MUSIC}}(\omega) = \mathbf{a}^T(\omega^{-1})\mathbf{Q}_e\mathbf{Q}_e^H \mathbf{a}(\omega).
\label{eq:rmusic_function}
\end{equation}
This function can be expressed as
\begin{equation}
f_{\text{R-MUSIC}}(\omega) = P_{\text{MUSIC}}^{-1}(\theta)=\sum_{m=0}^{N-1}\sum_{n=0}^{N-1}\omega^{n-m}C_{mn},
\label{eq:rmusic_function2}
\end{equation}
where $\mathbf{C}=\mathbf{Q}_e\mathbf{Q}_e^H.$ By substituting $l=n-m$, \eqref{eq:rmusic_function2} yields
\begin{equation}
f_{\text{R-MUSIC}}(\omega) = \sum_{l=-(N-1)}^{N-1}C_{l}\omega^l,
\label{eq:rmusic_function3}
\end{equation}
where $C_l = \sum_{n-m=l}C_{mn}$ is the sum of elements of $\mathbf{C}$ on the $l$-th diagonal. 

The function in  \eqref{eq:rmusic_function3} defines a polynomial of $2(N-1)$ zeros with dependencies, i.e., the roots come in pairs (if $z$ is a root then so does $1/z^*$). In the ideal case the magnitude of the roots would be unity, but in practice, since $\widetilde{\mathbf{Q}}_e$ is used instead of $\mathbf{Q}_e$, this does not hold. Since $z$ and $1/z^*$ have the same phase and reciprocal magnitude, one zero is within the unit circle and the other outside. By the definition of $z$, only the phase carries the desired information; thus, by choosing the $N-1$ roots within the unit circle and selecting the subset of the $K$ closest to the unit circle $\hat{z}_k,\ k=1,\dots,K,$ we can obtain the DoAs from
\begin{equation}
\hat{\theta}_k = -\arcsin \Big( \frac{\lambda}{2\pi d}\angle \hat{z}_k \Big),\ k=1,\dots,K.
\label{eq:rmusic_DoAs}
\end{equation}
R-MUSIC often provides better performance than MUSIC in the low-SNR regime.

\subsection{Compressed Sensing: mixed $\ell_{2,1}$-norm minimization}
\label{ssec:sparse_techniques}

A different approach to the DoA estimation task is via the use of compressed sensing (CS) techniques, which relies on the fact that the DoAs can be sparse in the spatial domain. According to this approach the continuous DoA domain can be replaced by a given set of grid points $\mathcal{G}=\{\phi_{1},\dots, \phi_{|\mathcal{G}|}\}$, where $|\mathcal{G}|$ with $|\mathcal{G}|\gg N$ is the grid size. This leads to the dictionary $\mathbf{A}_G = \mathbf{A} (\mathcal{G})=[\mathbf{a}(\phi_{1}),\dots, \mathbf{a}(\phi_{|\mathcal{G}|})]$ of size $N\times |\mathcal{G}|$. Collecting all data snapshots in a matrix form $\mathbf{Y}=[\mathbf{y}(1),\dots,\mathbf{y}(T)]$, problem \eqref{eq:signal_model} can be compactly expressed as:
\begin{equation}
\mathbf{Y} = \mathbf{A}_G \Saug + \Eaug,
\label{eq:signal_model_compact}
\end{equation}
where $\Saug=[\mathbf{\bar{s}}(1),\dots, \mathbf{\bar{s}}(T)]$ is a $|\mathcal{G}|\times T$ matrix whose column $\mathbf{\bar{s}}(t)$ is an augmented version of the source signal $\mathbf{s}(t)$ and is defined by:
\begin{equation}
\bar{s}_i(t)= \begin{cases} s_k(t),\ \text{if } \phi_i=\theta_k \\ 0,\ \ \ \ \ \ \text{otherwise,} \end{cases} i=1,\dots,|\mathcal{G}|,\ t=1,\dots,T.
\label{eq:signal_model_compact}
\end{equation}
Due to the grid mismatch problem \cite{5710590}, a quantization error $\mathbf{H}$ occurs that is added to the noise matrix, i.e., $\Eaug=\mathbf{E}+ \mathbf{H},$ where $\mathbf{E}=[\mathbf{e}(1),\dots, \mathbf{e}(T)].$ Hence, we can resort to convex optimization tools \cite{eldar2012compressed, boyd2004convex} for the DoA estimation. The basis pursuit denoising (BPDN) formulation is given by:
\begin{equation}
\min_{\Saug} \| \Saug \|_{2,1}\ \text{s.t.}\ \| \mathbf{Y} - \mathbf{A}_G \Saug \|_F\leq \eta,
\label{eq:BPDN1}
\end{equation}
where $\| \Saug \|_{2,1}=\sum_{i=1}^{|\mathcal{G}|} \|  \Saug_{i,:} \|_2$ is the mixed $\ell_{2,1}$-norm \cite{5290295}. 

The mixed norm has several interesting properties: a) it preserves sparsity and b) it demonstrates enhanced robustness (compared to greedy approaches based on the $\ell_0$ pseudo-norm). However, since in practice, the number of snapshots $T$ is very high directly solving \eqref{eq:BPDN1} is typically avoided. Instead, we can first apply a dimensionality reduction technique by using the singular value decomposition (SVD) of the $\mathbf{Y}$ matrix, i.e., $\mathbf{Y=ULV}^H.$ The reduced dimension $N\times R$ data, where $R=\RANK(\mathbf{Y})$, are then transformed to $\mathbf{Y}_{Dr}=\mathbf{YV}\mathbf{D}_R^T,$ where $\mathbf{D}_R=[\mathbf{I}_R\ \mathbf{0}].$ By letting $\Saug_{Dr}=\Saug \mathbf{VD}_R^T$ and $\Eaug_{Dr}=\Eaug \mathbf{VD}_R^T$,  \eqref{eq:signal_model_compact} yields
\begin{equation}
\mathbf{Y}_{Dr} = \mathbf{A}_G \Saug_{Dr} + \Eaug_{Dr}.
\label{eq:signal_model_compact_Dr}
\end{equation}
Thus, we can solve a similar problem in a space of reduced dimensionality, expressed as
\begin{equation}
\min_{\Saug_{Dr}} \| \Saug_{Dr}\|_{2,1}\ \text{s.t.}\ \| \mathbf{Y}_{Dr} - \mathbf{A}_G \Saug_{Dr} \|_F\leq \eta,
\label{eq:BPDN2}
\end{equation}
and obtain the solution from $\Saug = \Saug_{Dr} \mathbf{D}_R\mathbf{V}^H.$ Finally, the DoAs are calculated by computing the power for each row of $\Saug$. The method, which is known as $\ell_{2,1}$-SVD, was introduced in \cite{5290295} and was also employed in \cite{7314978}. The solution of \eqref{eq:BPDN2} can be obtained by using standard convex optimization packages.

\subsection{Multi-layer Prerceptron (MLP)}
\label{ssec:DL_doa_est}

The multi-layer perceptron architecture that was introduced in \cite{8485631} addresses DoA estimation of two sources and also considers array imperfections. The framework consists of a multitask autoencoder that acts as a group of spatial filters followed by a series of paralllel multi-layer DNNs for spatial spectrum estimation. The network is trained at each individual SNR\footnote{\url{https://github.com/LiuzmNUDT/DNN-DOA}}. However, as we will present in Section \ref{sec:results}, the network is not always successful in resolving DoAs. Of course, this depends on a number of factors, such as the SNR (low or high), the number of snapshots, as well as the angular separation of the DoAs.

\section{A Deep Convolutional Neural Network for DoA Estimation}
\label{sec:DoA_est_CNN}

Next, we formulate the task of estimating the DoAs as a multilabel classification task. In particular, we train a deep CNN that learns to predict the DoAs. The convolution layers perform the feature extraction from the multi-channel input data, and, subsequently, the fully connected (FC) layers use the output of the convolution layers to infer the DoA estimates using a pre-selected grid. In Section \ref{ssec:data_management}, we present the data management and labeling approach, whereas Section \ref{ssec:prop_CNN} is devoted to the description of the architecture.

\subsection{Data Management and Labeling}
\label{ssec:data_management}

DoA prediction is modeled as a \textit{multilabel classification task}. For $\phi_{\max} \in \{1^\circ,\dots,90^\circ\}$ we consider $2G+1$ discrete points of resolution $\rho$ (in degrees), which define a grid $\mathcal{G} = \{-G\rho, \dots, -\rho,0^\circ, \rho,\dots, G\rho \}\subset [-90^\circ,90^\circ]$, such that $\phi_{\max}= G\rho$. At each SNR level, $K$ angles are selected from the set $\mathcal{G}$ and the respective covariance matrix is calculated according to \eqref{eq:theor_cov_matrix}. The input data $\mathbf{X}$ to the proposed CNN is a real-valued $N\times N \times 3$ matrix, whose third dimension represents different ``channels.'' In particular, the first and second channels are the real and imaginary parts of $\mathbf{R}_y$, i.e., $\mathbf{X}_{:,:,1}=\Re \{\mathbf{R}_y \}$ and $\mathbf{X}_{:,:,2}=\Im \{\mathbf{R}_y\}$, whereas the third channel corresponds to phase entries, i.e., $\mathbf{X}_{:,:,3}=\angle \{\mathbf{R}_y\}$. Thus, the input data to the CNN is a collection of $D$ data points defined as $\mathcal{X} = \{\mathbf{X}_{(1)},\dots,\mathbf{X}_{(D)}\}$.

Next, for each example $\mathbf{X}_{(i)}$, the $K$ training angles in $\mathcal{G}$ are transformed into a binary vector with $K$ ones (the rest are zeros). For example, if $\phi_{\max}=60^\circ$ and the desired resolution is $\rho=1^\circ$, the grid becomes $\mathcal{G} = \{-60^\circ, \dots, -1^\circ,0^\circ, 1^\circ,\dots, 60^\circ \}$ with $121$ grid points. Moreover, the angle pair $\{-60^\circ,-59^\circ\}$ corresponds to the $121\times 1$ binary vector $\mathbf{z}=[1,1,0, \dots,0]^T,$ which serves as the corresponding label/output of the proposed CNN. Thus, the $i$-th label $\mathbf{z}_{(i)}$ belongs to the set $\mathcal{Z}=\{0,1\}^{2G+1}$ according to the described process. Hence, the $i$-th training example consists of pairs in the form $(\mathbf{X}_{(i)},\mathbf{z}_{(i)})$ leading to the training data set $\mathcal{D}=\{(\mathbf{X}_{(1)},\mathbf{z}_{(1)}),(\mathbf{X}_{(2)},\mathbf{z}_{(2)}),\dots,(\mathbf{X}_{(D)},\mathbf{z}_{(D)})\}$ of size $D$.

According to the well-known universal approximation theorem \cite{hornik1989multilayer} a feed-forward network with a single hidden layer processed by a multilayer perceptron can approximate continuous functions on a compact subset of $\R^n$. The goal of this multilabel classification task is to induce a ML hypothesis defined as a function $f$ from the input space to the output space, i.e., $f: \R^{N \times N \times 3} \rightarrow \mathcal{Z}.$ Although the true covariance matrix is used for training the network, for its testing and evaluation the sample covariance in \eqref{eq:sam_cov_matrix} is used, since the former is unknown. To this end, during the testing phase of the CNN all input examples can be considered as ``unseen data'' to the training.

\subsection{The Proposed CNN's Architecture}
\label{ssec:prop_CNN}

The nonlinear function $f$ is parametrized by a CNN of $24$ layers in total\footnote{Not all layers have trainable parameters.}, i.e.,
\begin{equation}
f(\mathbf{X}) = f_{24}(f_{23}(\dots f_1(\mathbf{X})))=\mathbf{z}.
\label{eq:layer_mapping}
\end{equation}

A standard CNN architecture was employed \cite{726791, gu2018recent} with only a few modifications that need to be taken into account due to the different nature of the problem we address. The functions $\{f_i(\cdot) \}_{i=1,4,7,10}$ represent the 2-dimensional (2D) convolutional layers of the network, whose number of filters is $n_C=256$. The kernel size is $\kappa \times \kappa$ and for the first convolutional layer $\kappa=3$, whereas $\kappa=2$ for the rest of them. The stride that we used is $\delta \times \delta$, where $\delta=1$ for all convolutional layers (and no padding) except for the first one, where $\delta=2$. Hence, for the input data $\mathbf{X} \in \R^{N\times N \times 3}$ and filter $\mathbf{K} \in \R^{\kappa \times \kappa \times 3}$ the mathematical expression of the convolution operation is a 2D matrix given by: 
\begin{equation}
(\mathbf{X}*\mathbf{K})_{m,n} = \sum_{i=1}^N \sum_{j=1}^N \sum_{k=1}^3 K_{i,j,k} X_{m+i-1,n+j-1,k},
\label{eq:Conv_2D_oper}
\end{equation}
whose dimension is $\lfloor (N-\kappa)/\delta+1 \rfloor \times \lfloor (N-\kappa)/\delta+1 \rfloor$. Thus, at the $\ell$-th layer: 
\begin{itemize}
\item The number of filters is denoted as $n_C^{[\ell]}$ and each kernel $\mathbf{K}^{(q)}$ has dimension $\kappa^{[\ell]}\times\kappa^{[\ell]}\times n_C^{[\ell-1]}$ for every $q=1,2,\dots,n_C^{[\ell]}$ with $n_C^{[0]}=3$;
\item $\delta^{[\ell]}$ is the stride;
\item $\mathbf{X}^{[\ell-1]}$ denotes the input of size $N^{[\ell-1]}\times N^{[\ell-1]} \times n_C^{[\ell -1]}$ with $\mathbf{X}^{[0]}=\mathbf{X}$ and $N^{[0]}=N$;
\item The bias of the $q$-th convolution is denoted as $b_q^{[\ell]}$;
\item $\mathbf{X}^{[\ell]}$ is the output of the convolutional layer and its size is $N^{[\ell]}\times N^{[\ell]} \times n_C^{[\ell]}$.
\end{itemize}
The convolution operation at the $\ell$-th layer for every $q=1,2,\dots,n_C^{[\ell]}$ is expressed as:
\begin{align}
(\mathbf{X}^{[\ell-1]}&*\mathbf{K}^{(q)})_{m,n} = \nonumber \\
& \sum_{i=1}^{N^{[\ell-1]}} \sum_{j=1}^{N^{[\ell-1]}} \sum_{k=1}^{n_C^{[\ell-1]}} K_{i,j,k}^{(q)} X_{m+i-1,n+j-1,k}^{[\ell-1]} + b_q^{[\ell]},
\label{eq:Conv_2D_layer}
\end{align}
whose dimension is $N^{[\ell]}\times N^{[\ell]}.$ Thus, the learned parameters at the $\ell$-th layer are $(\kappa^{[\ell]}\times \kappa^{[\ell]} \times n_C^{[\ell-1]})\times n_C^{[\ell]}$ for the filters plus $n_C^{[\ell]}$ for the biases. Pooling layers were not used, since the loss of information resulted in poor performance.

\begin{figure*}[t]
\centering
\includegraphics[scale = 0.165]{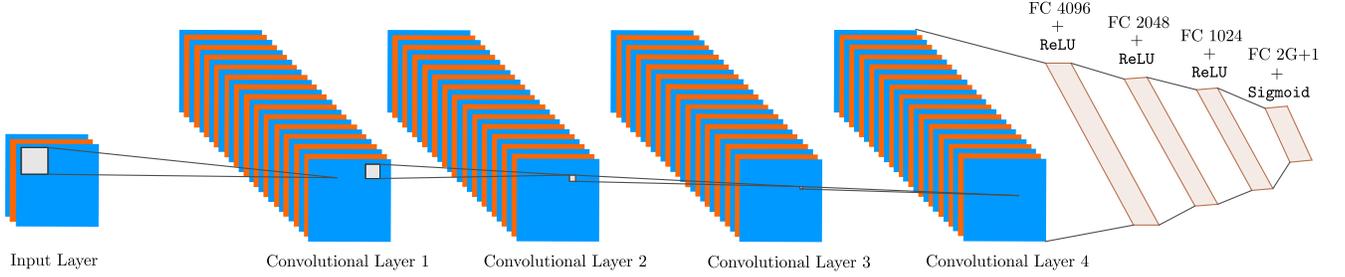}
\caption{The layout of the  proposed Convolutional Neural Network (CNN), which is used for the prediction of the DoAs via a supervised learning approach. Each convolutional layer has 256 filters and is followed by a batch normalization layer and a \texttt{ReLU} layer. After the Convolutional layers a flatten layer is used that transfers the learning process to fully connected (FC) layers, followed by a nonlinear activation function, which leads to the classification probability induced by the \texttt{Sigmoid} activation function at the final layer. Dropout layers are also employed (for regularization) in the first three FC layers.}
\label{fig:CNN_architecture}
\end{figure*}

Next, the functions $\{f_i(\cdot) \}_{i=2,5,8,11}$, which represent batch normalization layers are used. The following set of functions denoted as $\{f_i(\cdot) \}_{i=3,6,9,12}$ correspond to rectified linear unit (\texttt{ReLU}) layers, i.e., layers where the nonlinear function \texttt{ReLU} with \texttt{ReLU}$(x)=\max (0,x)$ (applied element wise). The $13$-th layer is a flatten layer with no trainable parameters, which shapes the tensor-valued output of the final convolutional layer to a vector. Thereafter, the fully connected (FC) layers follow. The functions $\{f_i(\cdot) \}_{i=14,17,20,23}$ are FC (dense) layers with 4096, 2048, 1024 and $2G+1$ neurons, respectively. Each $\ell$-th FC layer maps its input $\mathbf{c}^{[\ell-1]} \in \R^{M^{[\ell-1]}}$ to the output $\mathbf{c}^{[\ell]} \in \R^{M^{[\ell]}}$ via a set of weights $\mathbf{W}^{[\ell]} \in \R^{M^{[\ell]}\times M^{[\ell-1]}}$ and biases $\mathbf{b}_{\text{FC}}^{[\ell]}\in \R^{M^{[\ell]}}.$ Thus, the output of the $\ell$-th FC layer is given by
\begin{equation}
\mathbf{c}^{[\ell]} = \mathbf{W}^{[\ell]}\mathbf{c}^{[\ell-1]}+\mathbf{b}_{\text{FC}}^{[\ell]},
\label{eq:FC_layers}
\end{equation}
where the set of parameters $\vartheta^{[\ell]}=\{\mathbf{W}^{[\ell]}, \mathbf{b}_{\text{FC}}^{[\ell]}\}$ with $M^{[\ell-1]}\times M^{[\ell]}+ M^{[\ell]}$ entries (in total) is optimized during the training of the neural network. After each FC layer follows a nonlinear activation function (applied element-wise). The functions $\{f_i(\cdot) \}_{i=15,18,21}$ correspond to \texttt{ReLU} layers, whereas $\{f_i(\cdot) \}_{i=16,19,22}$ correspond to Dropout layers that randomly set weights to zero with probability $20\%$ (non trainable parameters). For the final (output) layer, i.e., $f_{24}(\cdot)$, we utilize the \texttt{Sigmoid} function $s(x)=e^x/(e^x+1)$ with a return value in $[0,1].$ The selection of the sigmoid function over the softmax is due to the presence of $K$ labels, which could independently receive a value equal (during the training) or close (during the inference) to 1. Thus, the output of the CNN is a probability at each entry of the predicted label, which for the input data $\mathbf{X}_{(i)}$ is expressed as 
\begin{equation}
\hat{\mathbf{p}}_{(i)}=f(\mathbf{X}_{(i)})=\begin{pmatrix}
\hat{p}_1\\ \vdots \\ \hat{p}_{2G+1}
\end{pmatrix}.
\label{eq:predictor}
\end{equation}
The layout of the proposed CNN is depicted in Fig. \ref{fig:CNN_architecture}.

The training of the CNN is performed in a supervised manner over the training data set $\mathcal{D}$. In particular, since the adopted approach is a multilabel classification task, we attempt to optimize the set of all trainable parameters $\vartheta$ whose updates are carried out via back-propagation by minimizing the reconstruction error, i.e.:
\begin{equation}
\vartheta^* = \argmin_{\vartheta} \frac{1}{D} \sum_{i=1}^{D} \mathrm{L}\left( \hat{\mathbf{p}}_{(i)}; \mathbf{z}_{(i)} \right),
\label{eq:CNN_min}
\end{equation}
where 
\begin{align}
\label{eq:CNN_loss}
\mathrm{L}(\hat{\mathbf{p}}_{(i)}; \mathbf{z}_{(i)}) &=   - \sum_{n=1}^{2G+1} [z_{(i)}(n)\log (\hat{p}_{(i)}(n)) \nonumber \\
&\quad + (1 - z_{(i)}(n))\log (1-\hat{p}_{(i)}(n))],
\end{align}
is the binary cross-entropy loss.

\section{Training Approach}
\label{sec:training_approach}

For the training of the proposed CNN we considered a grid of $\rho =1^\circ$ resolution and $\phi_{\max} =60^\circ$ (corresponding to the grid set $\mathcal{G}=\{-60^\circ,\dots, 60^\circ\}$) with $121$ grid points, which is the output dimension of the CNN after the binary transformation of the DoAs. We resorted to data from the true covariance matrix, hence, a) reducing the number of training data required and b) enabling the DoA prediction from any sufficient collection of snapshots for the sample covariance estimate. One of the main advantages of the proposed approach is that the CNN can also be trained to infer the number of sources, since the problem is modeled as a multi-classification task. Therefore, we adopt two approaches for the training of the CNN: in Section \ref{ssec:scenario1} we consider \emph{a fixed number of sources}, which is more suitable in cases where the number of sources is known a priori. In Section \ref{ssec:scenario2}, the network is trained for a varying number of sources; therefore, we will refer to the second one as a \emph{mixed number of sources} training approach, which is more suitable in cases where \emph{the number of sources/targets is unknown}, such as in military applications. The experimental section, where the proposed model is evaluated, is also separated into two parts, according to the training options described next.

\subsection{Fixed Number of Sources}
\label{ssec:scenario1}

The number of sources is set to $K=2$. For each SNR level the training set consists of all the $\binom{2G+1}{K}$ combinations\footnote{The order of the DoAs does not play any role, since the covariance matrix (input to the CNN) remains the same for the given angles.}, which leads to a training data size of $7,260$ examples per SNR level. For training with a fixed number of sources, we used the examples from all SNRs jointly. We also observed that training in the low-SNR regime (worst case scenario training approach) is also sufficient for prediction at higher SNRs. To this end, we trained the proposed CNN across a range of low-SNRs, i.e., at SNRs ranging from $-20$ to $0$ dB with an increasing step of 5. Thus, this led to $D=5 \cdot 7,260=36,300$ training examples in total. Once the CNN is trained prediction can be performed at higher-SNRs, as we experimentally demonstrate in Section \ref{ssec:knownK}. Finally, we also observed that training at each individual SNR could slightly improve the performance of the proposed CNN; however, considering the required effort (training that many networks and storing the trained parameters) with respect to the insignificant gains, we have decided to follow the described joint SNR training approach.

In the training phase, the data were randomly split into training ($90\%$) and validation sets ($10\%$). For the update/optimization of the CNN's parameters we employed Adam \cite{kingma2014adam} with an initial learning rate of $0.001$, $\beta_1=0.9$, $\beta_2=0.999$, but we reduced the learning rate by $50\%$ every 10 epochs in order to guarantee convergence to a solution. The batch size was set to 32 and the network was trained for 200 epochs. The CNN was implemented in Keras using Tensorflow as backend; the operating system was Windows running on an Intel Xeon Gold 5222 processor at 3.8 GHz and with an NVIDIA TITAN RTX GPU.

\subsection{Mixed Number of Sources}
\label{ssec:scenario2}

Next, we describe our approach in order to train the proposed CNN for a varying number of sources, ranging from $1$ to $K_{max}$. In the numerical section we  selected $K_{max}=3$. At each SNR level the training set consists of $\sum_{k=1}^{K_{max}}\binom{2G+1}{k}=295,361$ training examples (summation of all the combinations of angles) for $G=60$. Each training label is an $121\times 1$ binary vector with one to $K_{max}=3$ 1's and the rest of its entries equal to zero. The fact that $K$ now varies makes the learning process more demanding; therefore, we opted for the training at the individual SNRs. Nevertheless, joint training for a range of SNRs is still an option (as discussed in Section \ref{ssec:scenario1}) with the advantage of obtaining a single set of optimized parameters at the cost of a small compromise in performance. In particular, we trained the proposed CNN at $-10$ and at $0$ dB SNR. The network's training parameters are the same as the ones used for the fixed number of sources, except for the learning rate decrease cycle which is every 20 epochs.

\section{Simulation Results}
\label{sec:results}

In this section, we provide extensive simulation results, where we evaluate the performance of the proposed CNN in the DoA estimation task under various setups. Moreover, we compare the proposed approach to the MUSIC and R-MUSIC estimators, as well as the robust $\ell_{2,1}$-SVD method in \cite{7314978}. In the first part, we evaluate the performance of the CNN assuming that the number of sources is known a priori. In the second part, after training the network for a mixed number of sources (up to a maximum number), we relax the former assumption and provide results where the CNN is able to predict the number of sources jointly with the DoAs. 

In all experiments, we consider a ULA with $N=16$ elements equally space at half-wavelength distance ($d=\lambda/2$). Thus, the number of all trainable parameters of the proposed CNN is approximately $28$ million. For MUSIC and $\ell_{2,1}$-SVD the adopted grid resolution is chosen to be the same as that of the CNN ($\rho = 1^\circ$). On the other hand, R-MUSIC is a gridless estimator therefore it is expected to perform much better and optimally in the high-SNR regime. For all the experiments, the SNR is defined as in \cite{Wang2017a}:
\begin{equation}
\SNR = 10\log_{10} \frac{\min \{ \sigma_1^2,\sigma_2^2,\dots, \sigma_K^2 \} }{\sigma_{e}^2}.
\label{eq:SNR}
\end{equation}

\subsection{Number of Sources Known}
\label{ssec:knownK}

First, we consider the training strategy introduced in Section \ref{ssec:scenario1}. For each testing example $\X_{(i)}$ the output of the CNN is given by \eqref{eq:predictor}. Assuming that we know the number of sources, $K$, we can pick up the $K$ largest probabilities in \eqref{eq:predictor}. The estimated angles result from their corresponding grid points. 

The metric that is employed for the evaluation of the CNN is the empirical RMSE, which is defined as:
\begin{equation}
\RMSE = \sqrt{\frac{1}{D_{\text{te}} K} \sum_{k=1}^K \sum_{m=1}^{D_{\text{te}}} \left( \theta_k^{(m)}-\hat{\theta}_k^{(m)} \right)^2},
\label{eq:RMSE}
\end{equation}
where $[\theta_1^{(m)},\dots,\theta_K^{(m)}]^T$ are the actual DoAs (ground truth) and $[ \hat{\theta}_1^{(m)}, \dots,\hat{\theta}_K^{(m)}]^T$ are the estimated DoAs at the $m$-th testing example, whereas $D_{\text{te}}$ is the total number of testing examples for each experiment. 

\begin{figure*}
\centering
\subfloat[]{\includegraphics[width=0.25\textwidth]{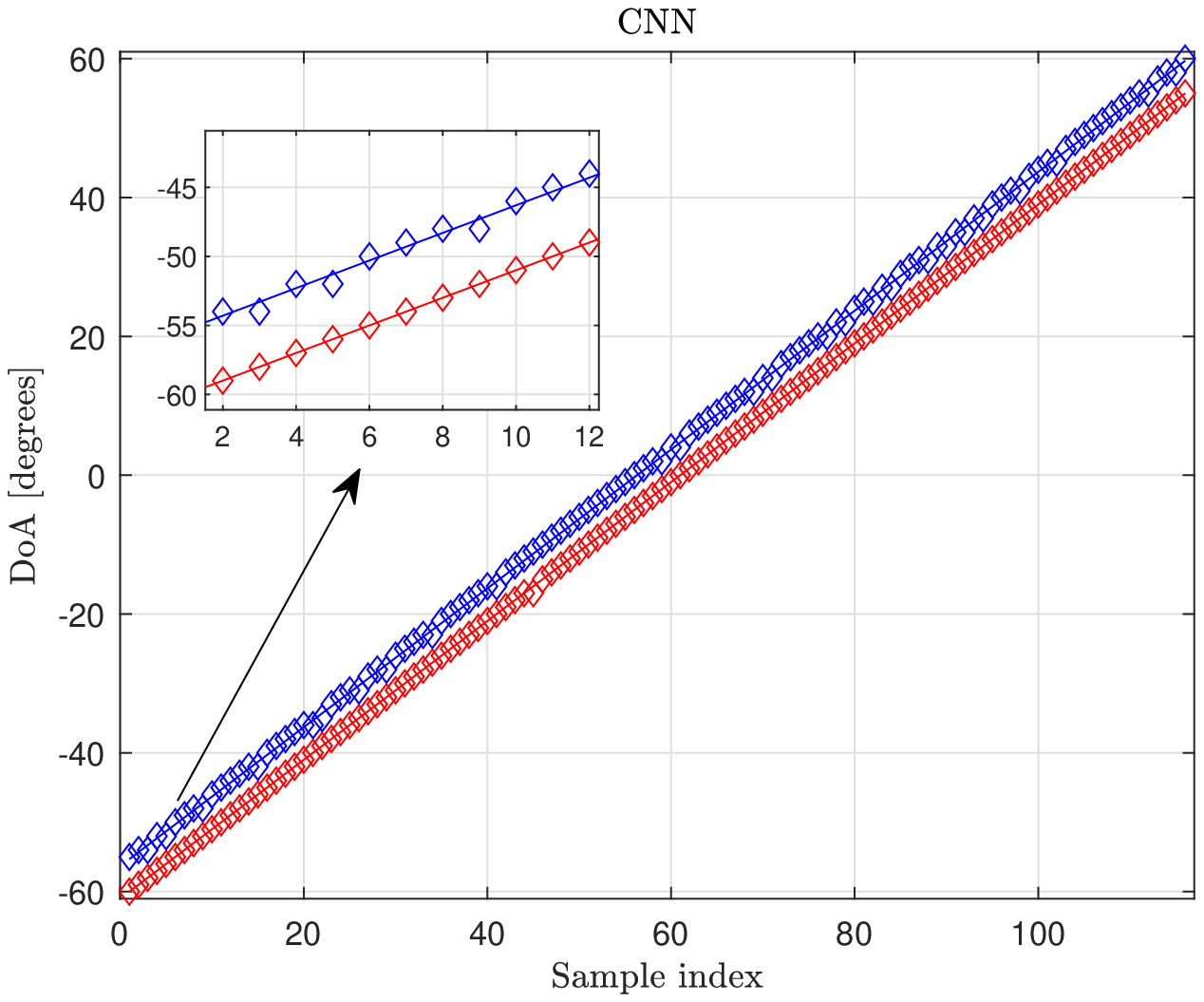}}\hfil
\subfloat[]{\includegraphics[width=0.25\textwidth]{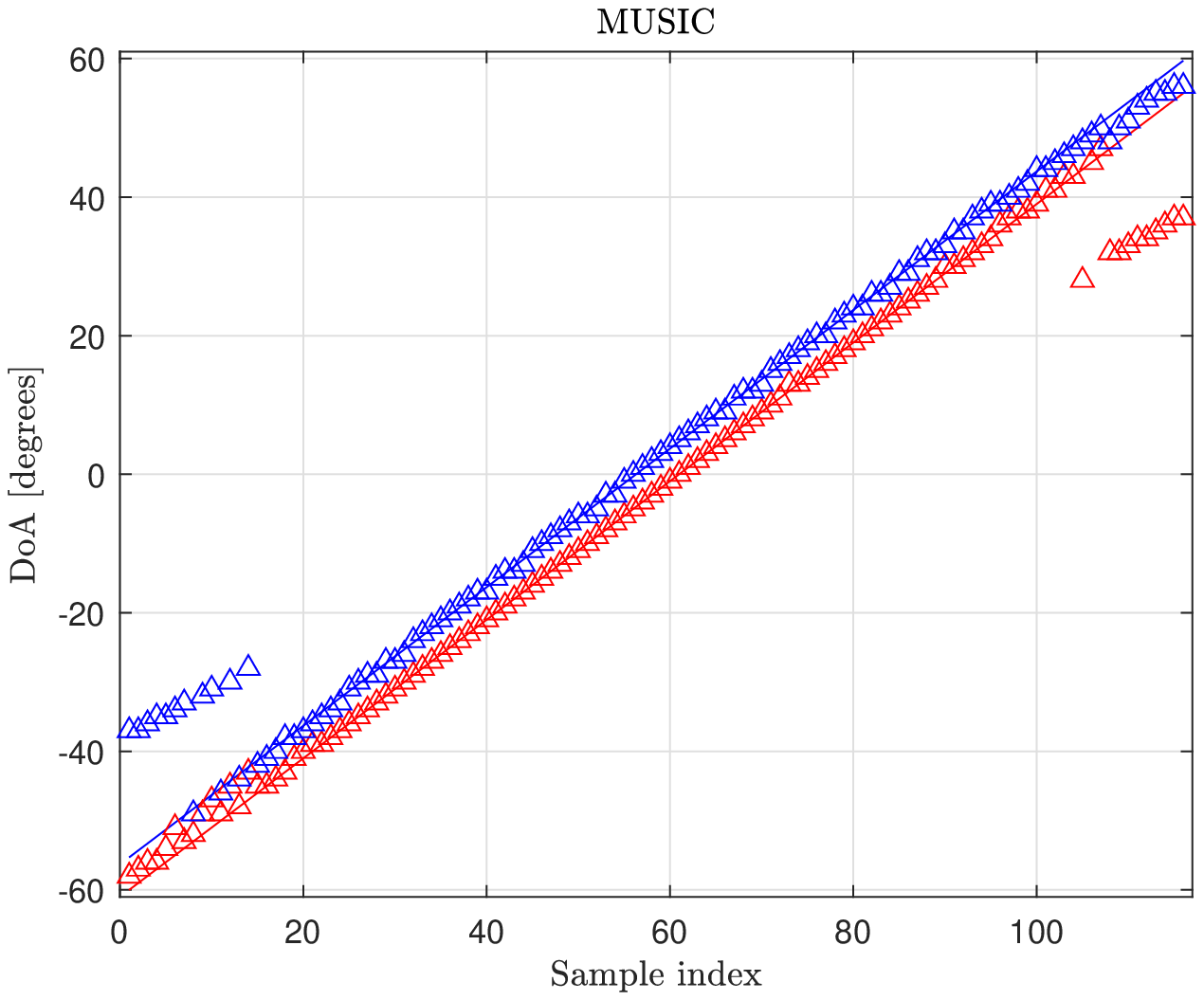}}\hfil 
\subfloat[]{\includegraphics[width=0.25\textwidth]{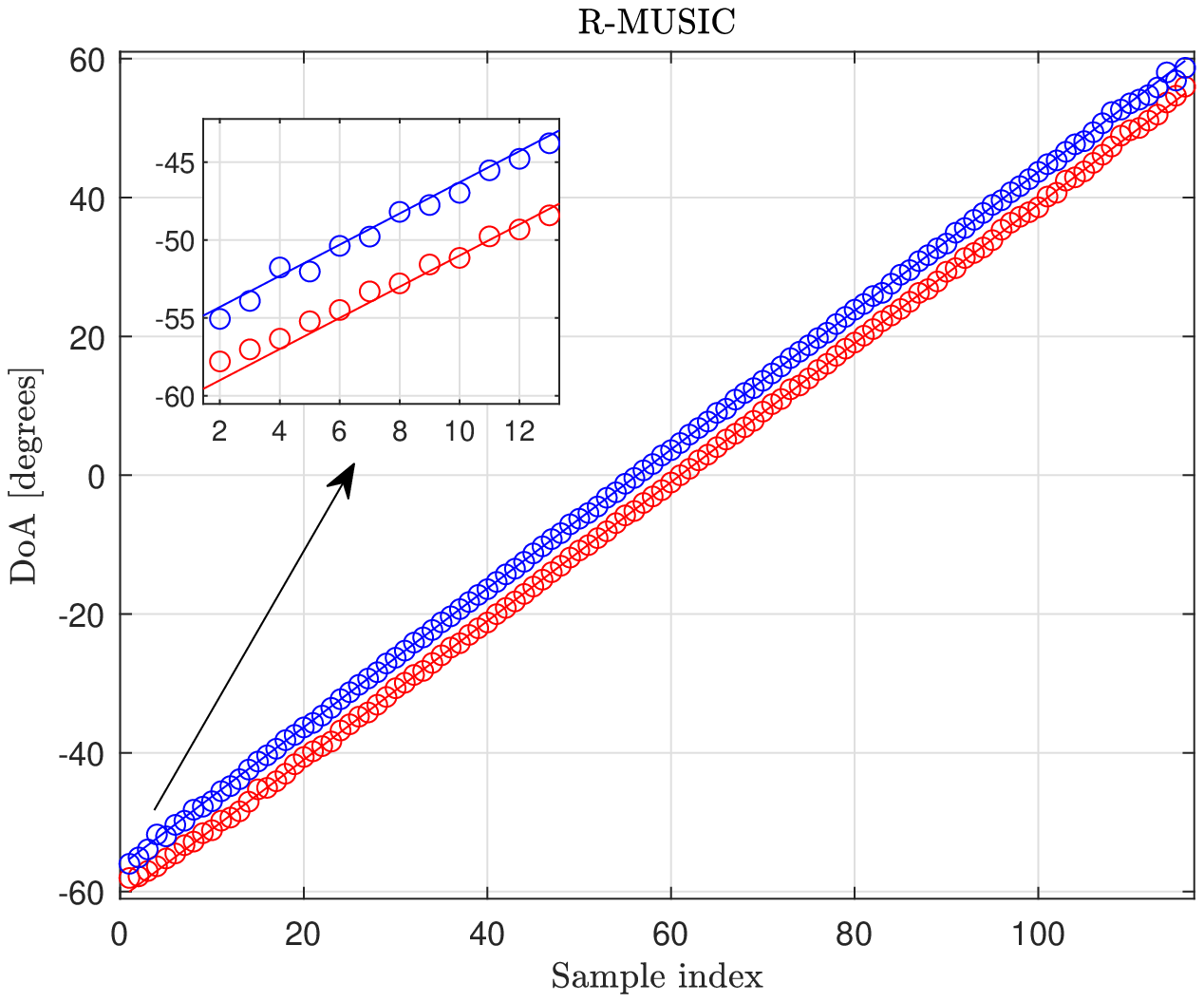}}\hfil
\subfloat[]{\includegraphics[width=0.25\textwidth]{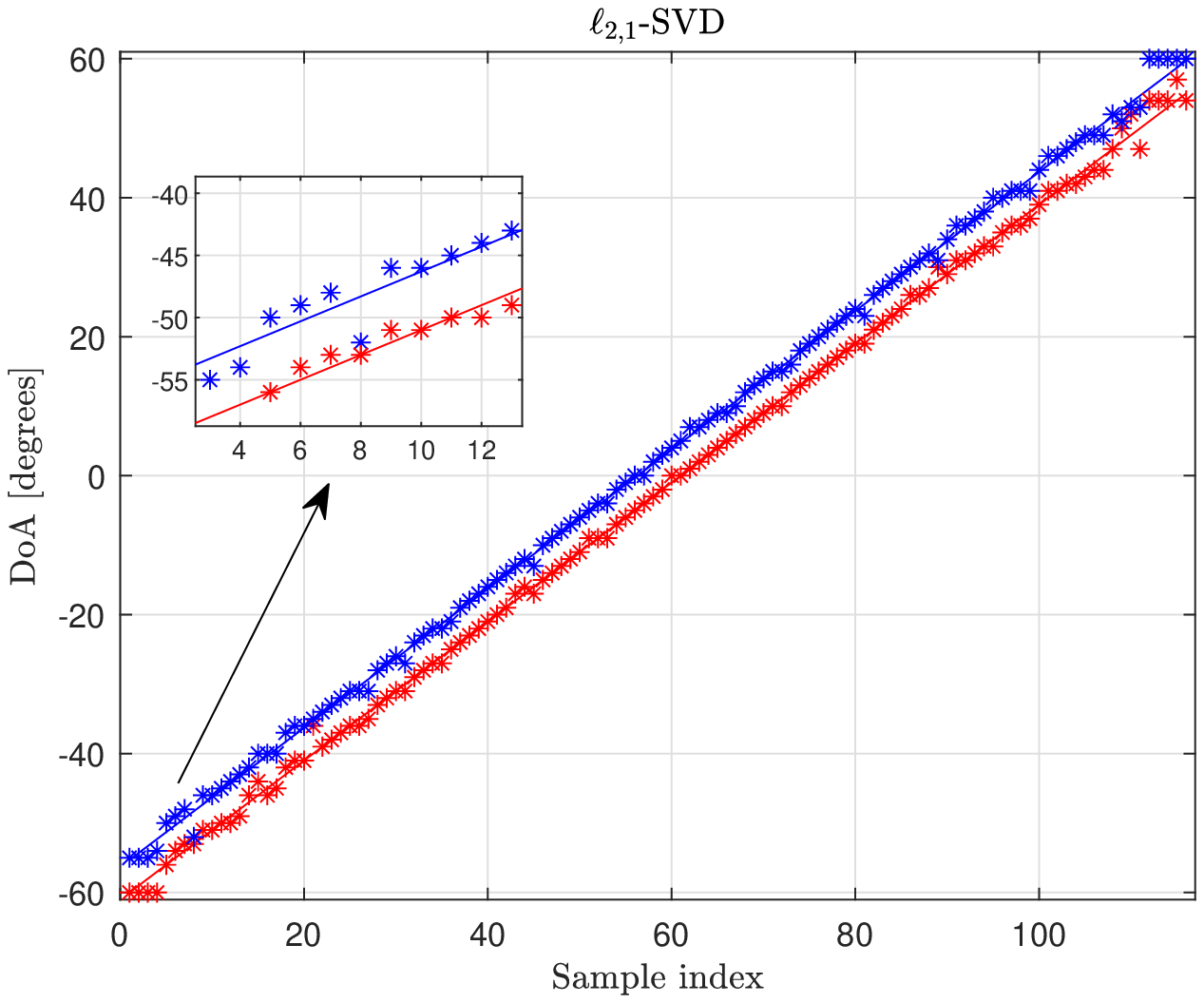}}

\subfloat[]{\includegraphics[width=0.25\textwidth]{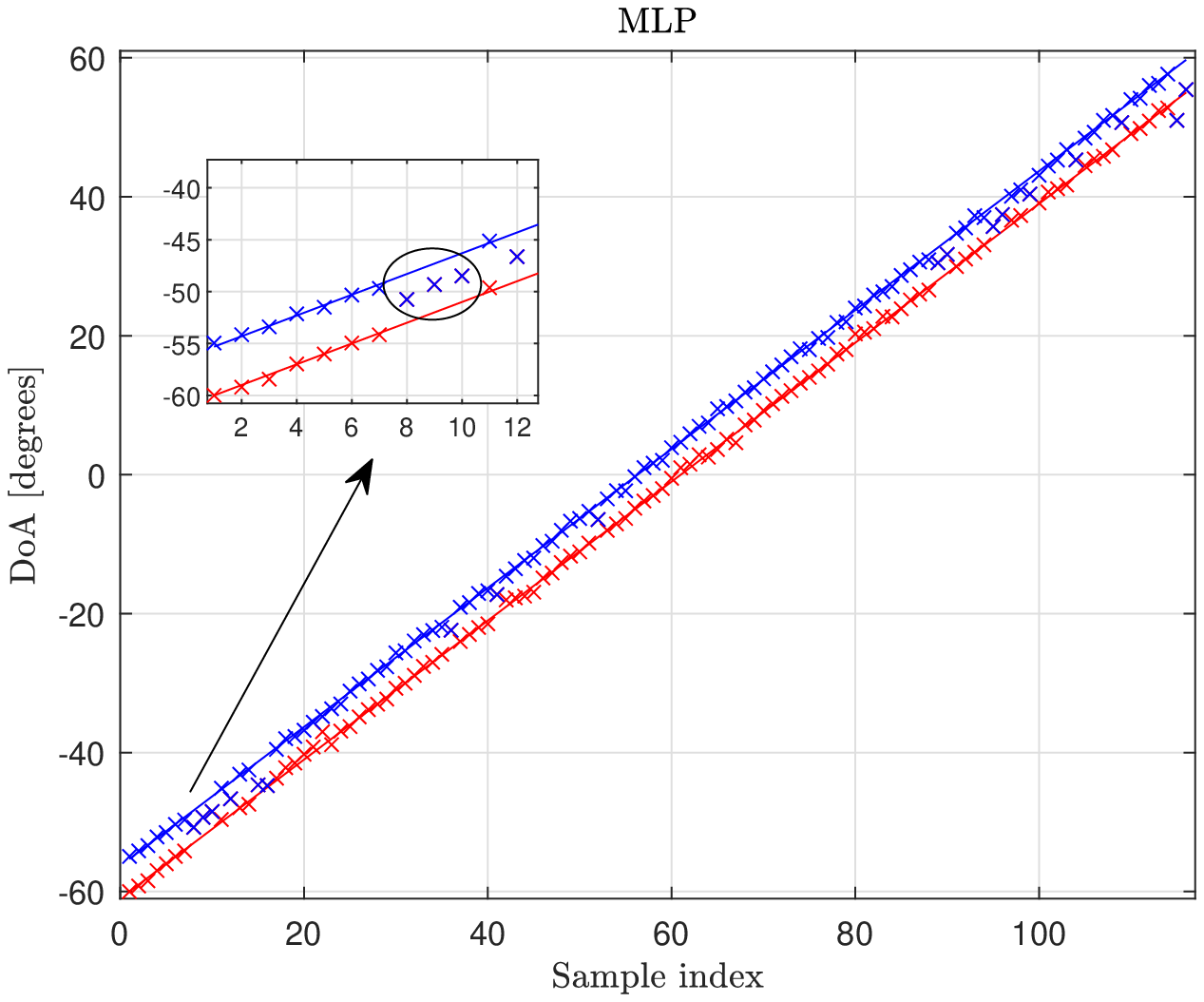}}\hfil   
\subfloat[]{\includegraphics[width=0.25\textwidth]{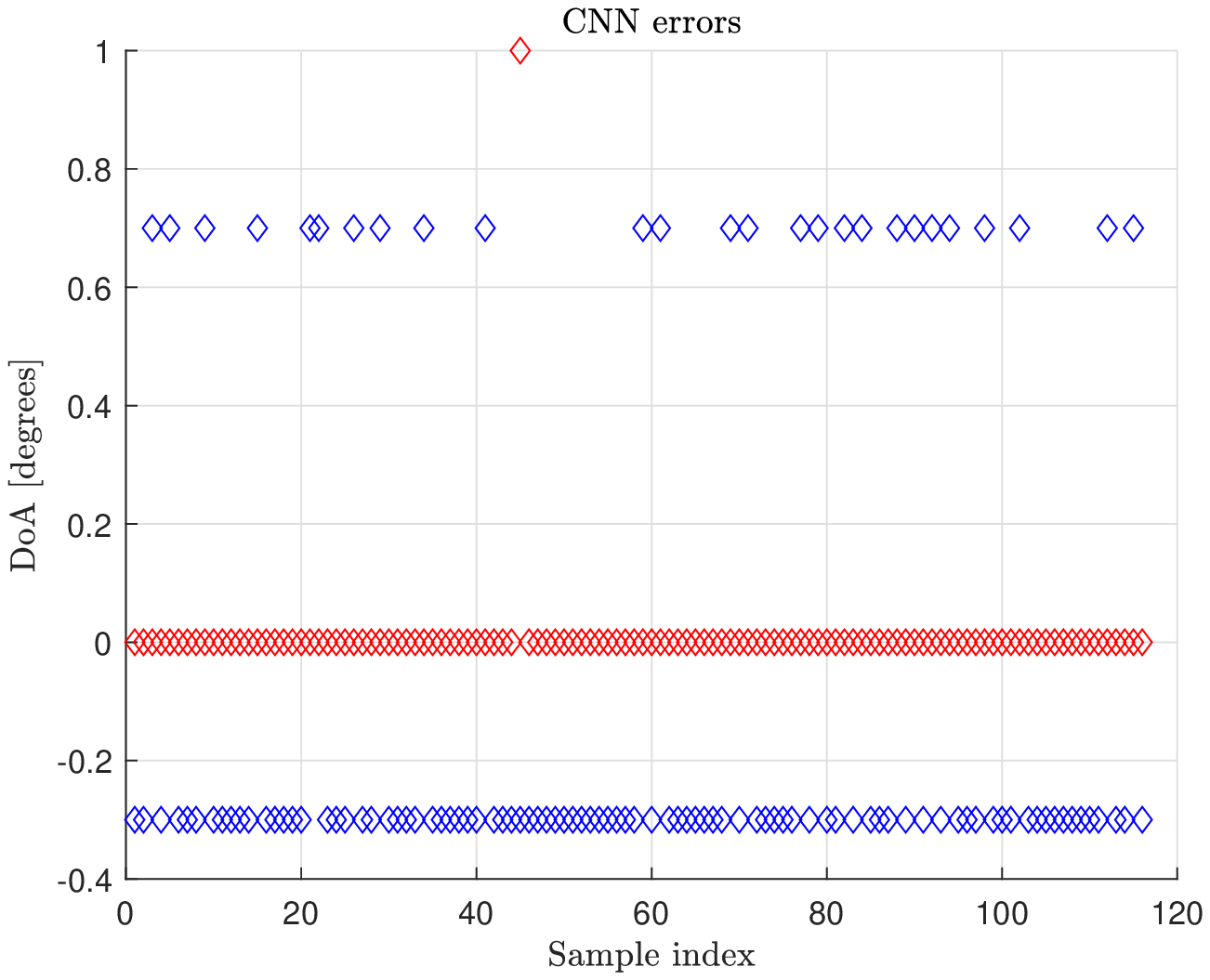}}\hfil
\subfloat[]{\includegraphics[width=0.25\textwidth]{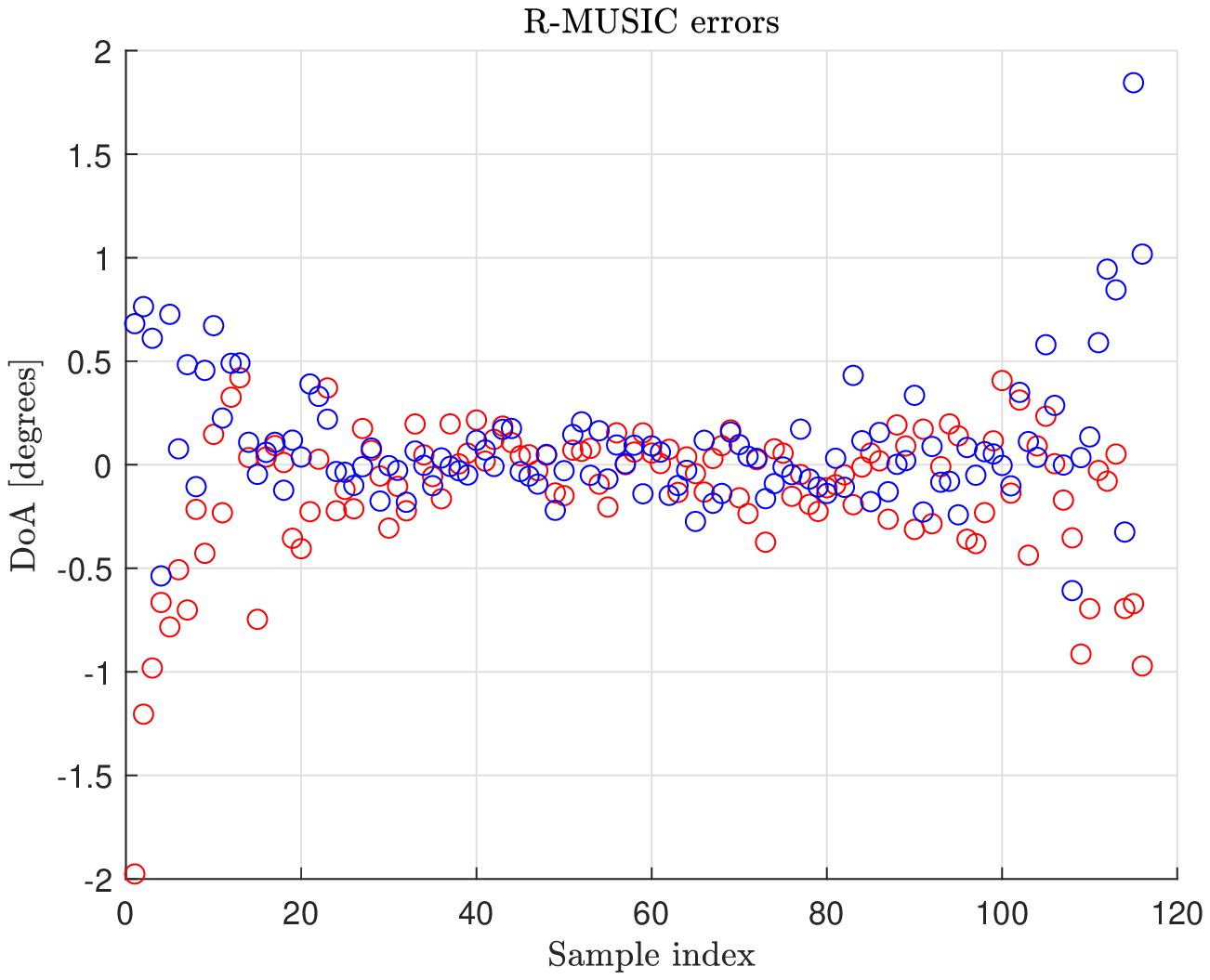}}

\caption{DoA estimation performance on off-grid angles at -10 dB SNR using $T=2,000$ snapshots. Estimated DoAs by the (a) CNN (proposed), (b) MUSIC, (c) R-MUSIC, (d) $\ell_{2,1}$-SVD and (e) MLP. (f) CNN's errors. (g) R-MUSIC's errors. The CNN outperforms its competitors in DoA estimation by i) attaining the smallest errors and ii) demonstrating a robust performance comparable to that of the grid-less R-MUSIC estimator.}
\label{fig:Exp1A_min10dB_SNR_offgrid_slideangles}
\end{figure*}

\subsubsection{Direction-of-Arrival Estimation and Errors}
\label{sssec:error_distr}

In the first set of experiments, we consider two DoA scenarios in the low-SNR regime. First, two signals with an angular distance of $\Delta \theta = 4.7^\circ$ at $\SNR=-10$ dB impinge onto the array, while the direction of the first signal varies from $-60^\circ$ to $55^\circ$ with an increasing step of $1^\circ$. The angular distance between the two sources is not contained in the training set, hence, the angular direction of the second signal deviates from those used for the training of the network. We considered that $T=2,000$ snapshots are collected for the sample covariance estimate. The DoAs predicted by the proposed CNN are depicted in Fig. \ref{fig:Exp1A_min10dB_SNR_offgrid_slideangles}(a) (the solid line corresponds to the actual DoAs), whereas in Fig. \ref{fig:Exp1A_min10dB_SNR_offgrid_slideangles}(f) the respective errors are plotted. Additionally, in Fig. \ref{fig:Exp1A_min10dB_SNR_offgrid_slideangles}(b) and (d), we have plotted the MUSIC and $\ell_{2,1}$-SVD DoA estimates, respectively. The DoA estimates of the MLP are shown in Fig. \ref{fig:Exp1A_min10dB_SNR_offgrid_slideangles}(e), whereas the estimated angles by the R-MUSIC are depicted in Fig. \ref{fig:Exp1A_min10dB_SNR_offgrid_slideangles}(c) with their respective errors in Fig. \ref{fig:Exp1A_min10dB_SNR_offgrid_slideangles}(g). We observe that, the majority of the CNN's errors lie in the interval $[-0.3^\circ,0.7^\circ]$ (with the exception of a single deviation of $1^\circ$), whereas the errors of the subspace-based methods are in $[-18.3^\circ,18^\circ]$ and $[-2^\circ,1.845^\circ]$ for the MUSIC and R-MUSIC estimators, respectively. Furthermore, the errors of the $\ell_{2,1}$-SVD lie in $[-4.3^\circ,3.7^\circ]$. Moreover, for the subspace-based methods we observe a higher density of errors located near the borders of the angular region, i.e., at $-60^\circ$ and $60^\circ$. This behavior does not occur with the CNN as we can see in Fig. \ref{fig:Exp1A_min10dB_SNR_offgrid_slideangles}(f). Overall, the empirical RMSE in the DoA estimation is  $0.31^\circ$ for the CNN, whereas $5.01^\circ,\ 1^\circ$ and $0.35^\circ$ for the MUSIC, the $\ell_{2,1}$-SVD and the R-MUSIC estimators, respectively. It should be noted that the MLP could not resolve DoAs in 18 out of the 116 examples, e.g., see the circled estimated DoAs in Fig. \ref{fig:Exp1A_min10dB_SNR_offgrid_slideangles}(e), in the current setup (providing two identical estimated directions). Therefore, the error of this estimator cannot be calculated. The threshold value used for the $\ell_{2,1}$-SVD method in this experiment is $\eta=550$.

\begin{figure*}
\centering
\subfloat[]{\includegraphics[width=0.25\textwidth]{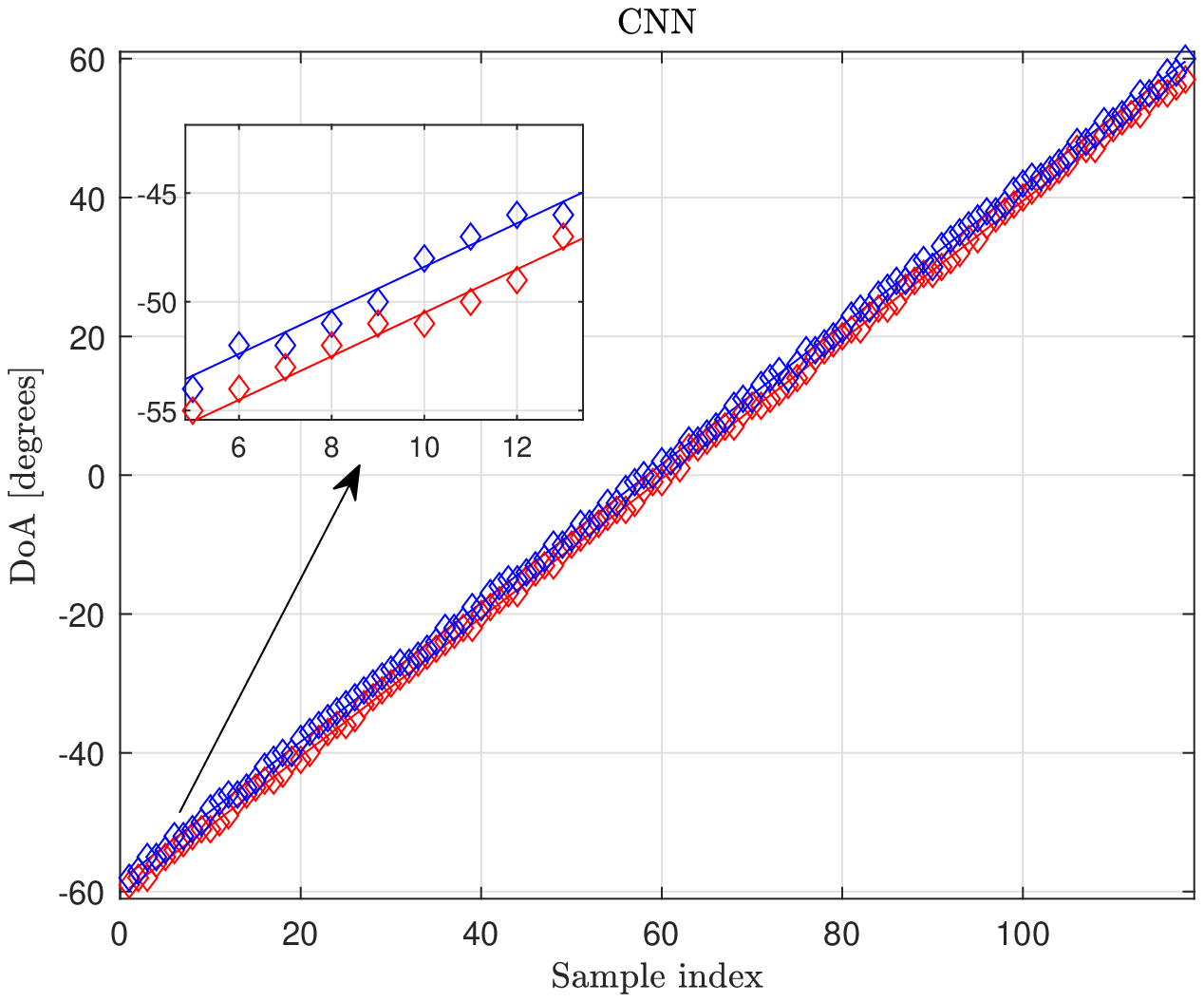}}\hfil
\subfloat[]{\includegraphics[width=0.25\textwidth]{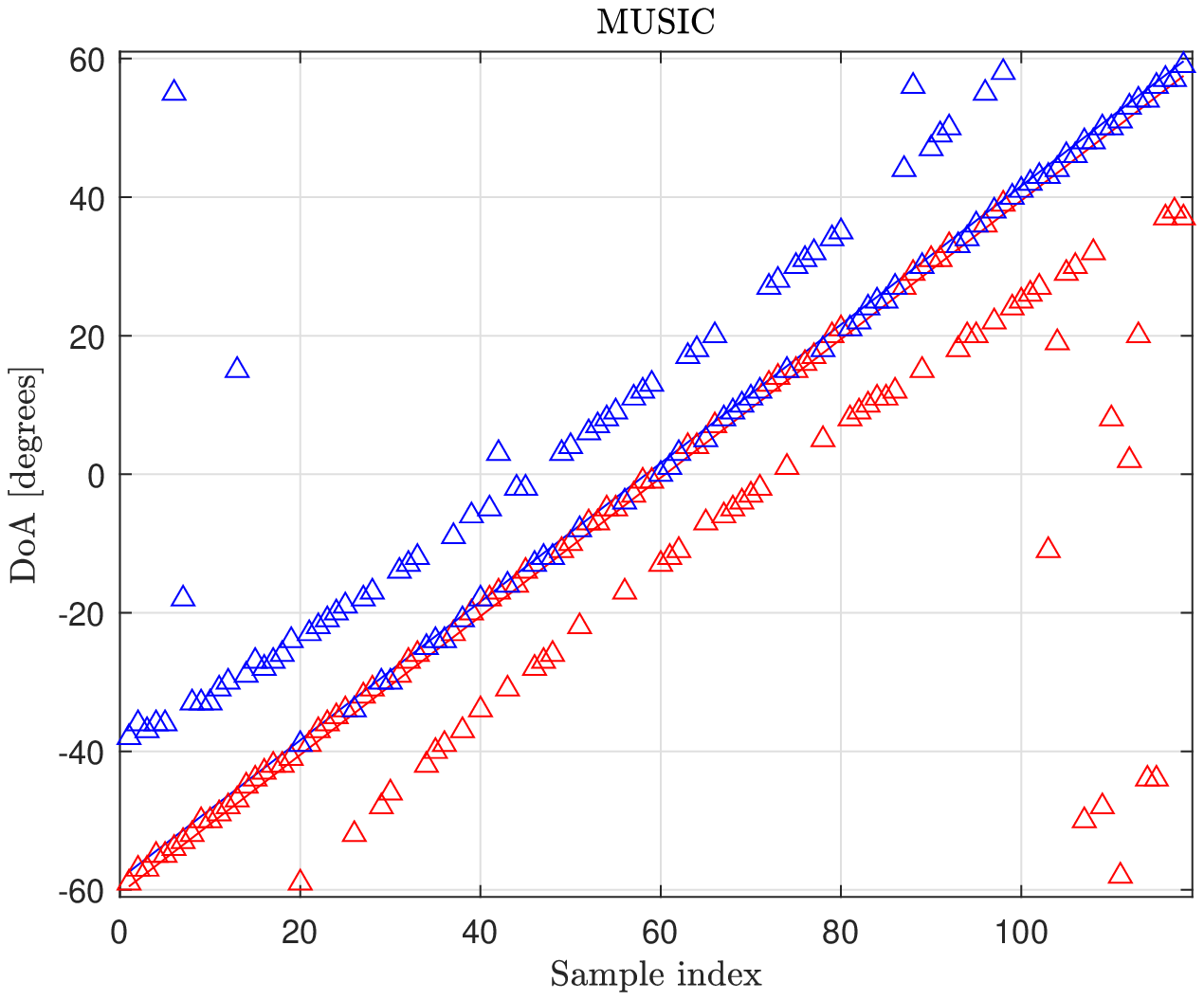}}\hfil 
\subfloat[]{\includegraphics[width=0.25\textwidth]{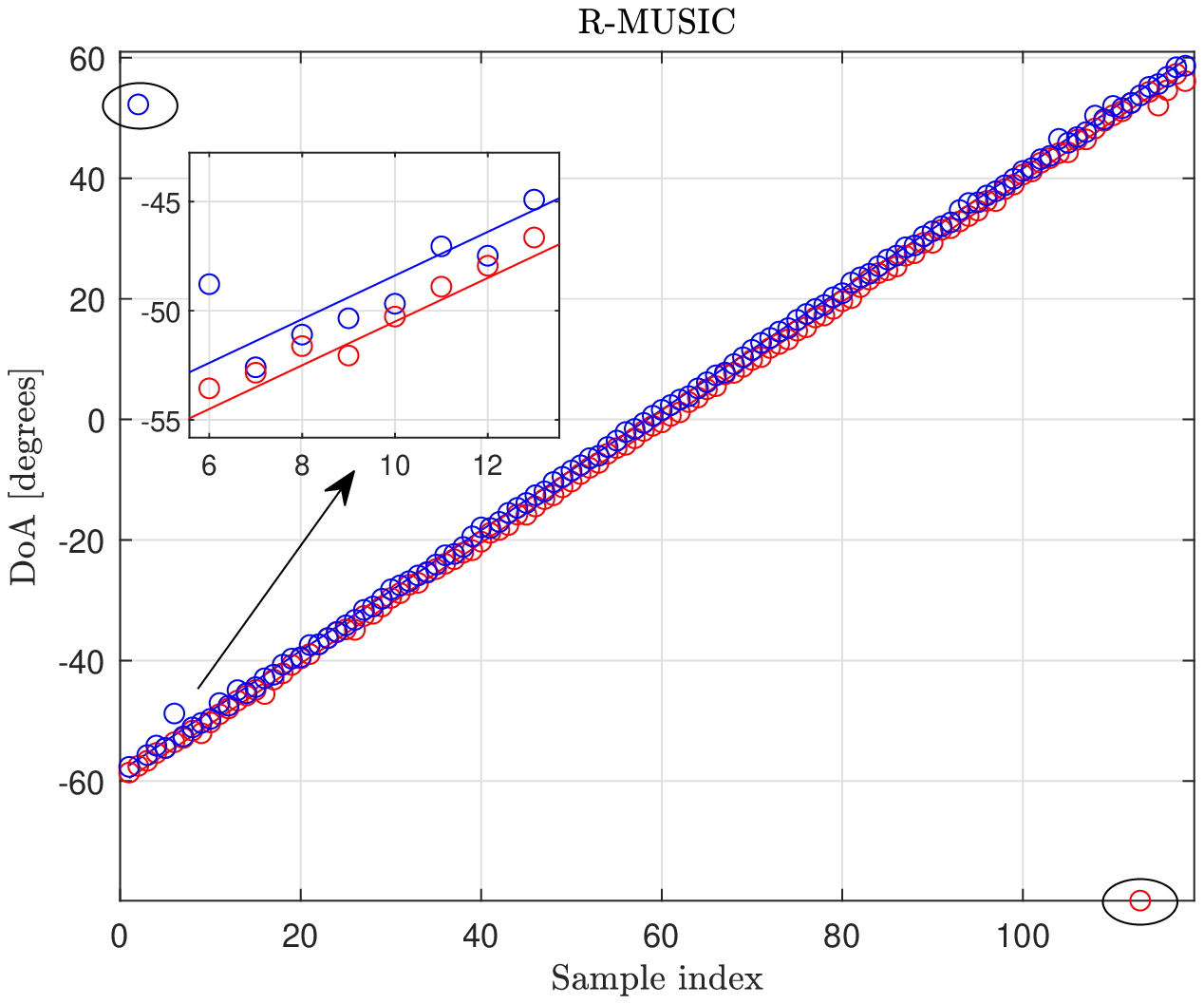}}\hfil 
\subfloat[]{\includegraphics[width=0.25\textwidth]{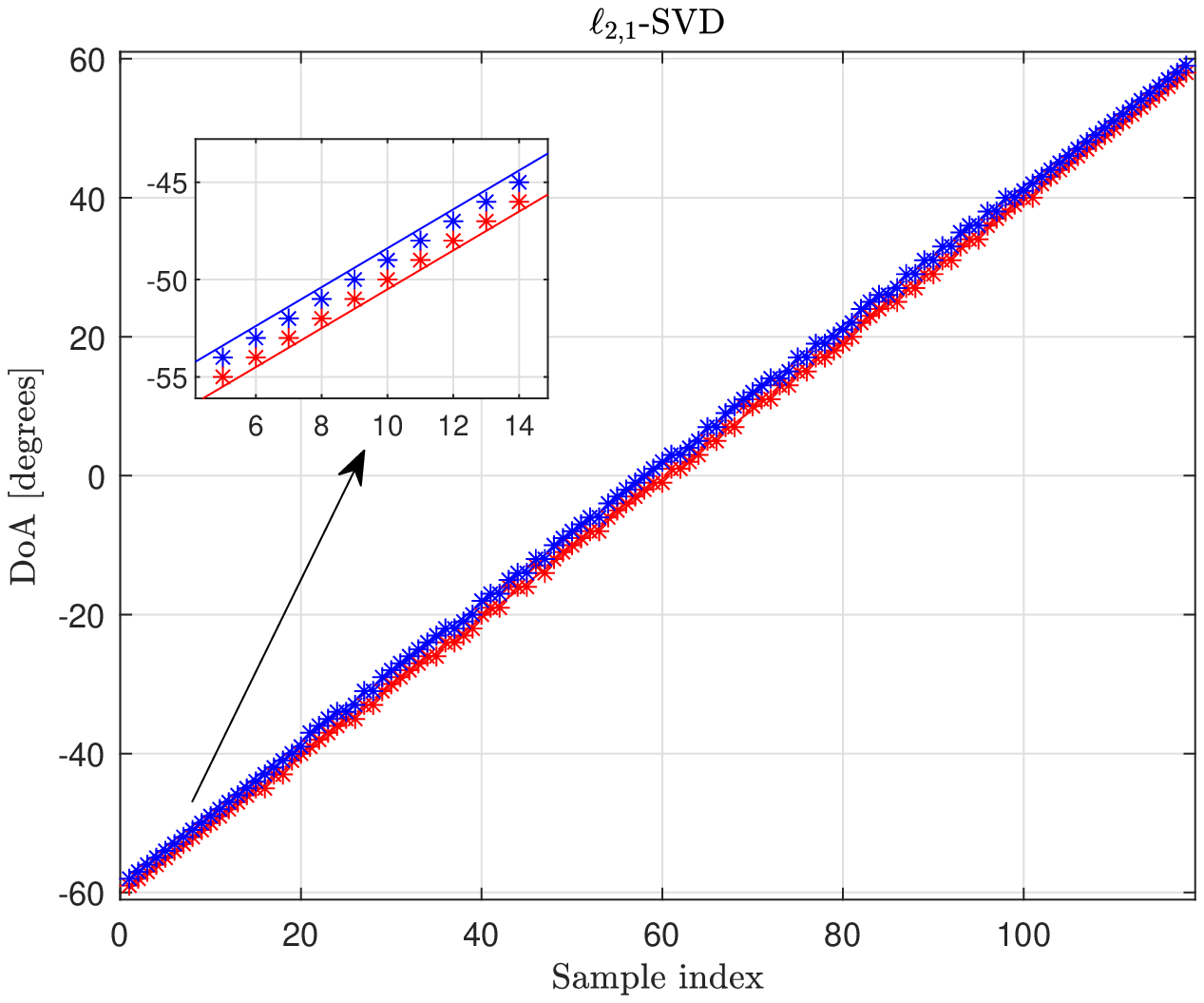}}

\subfloat[]{\includegraphics[width=0.25\textwidth]{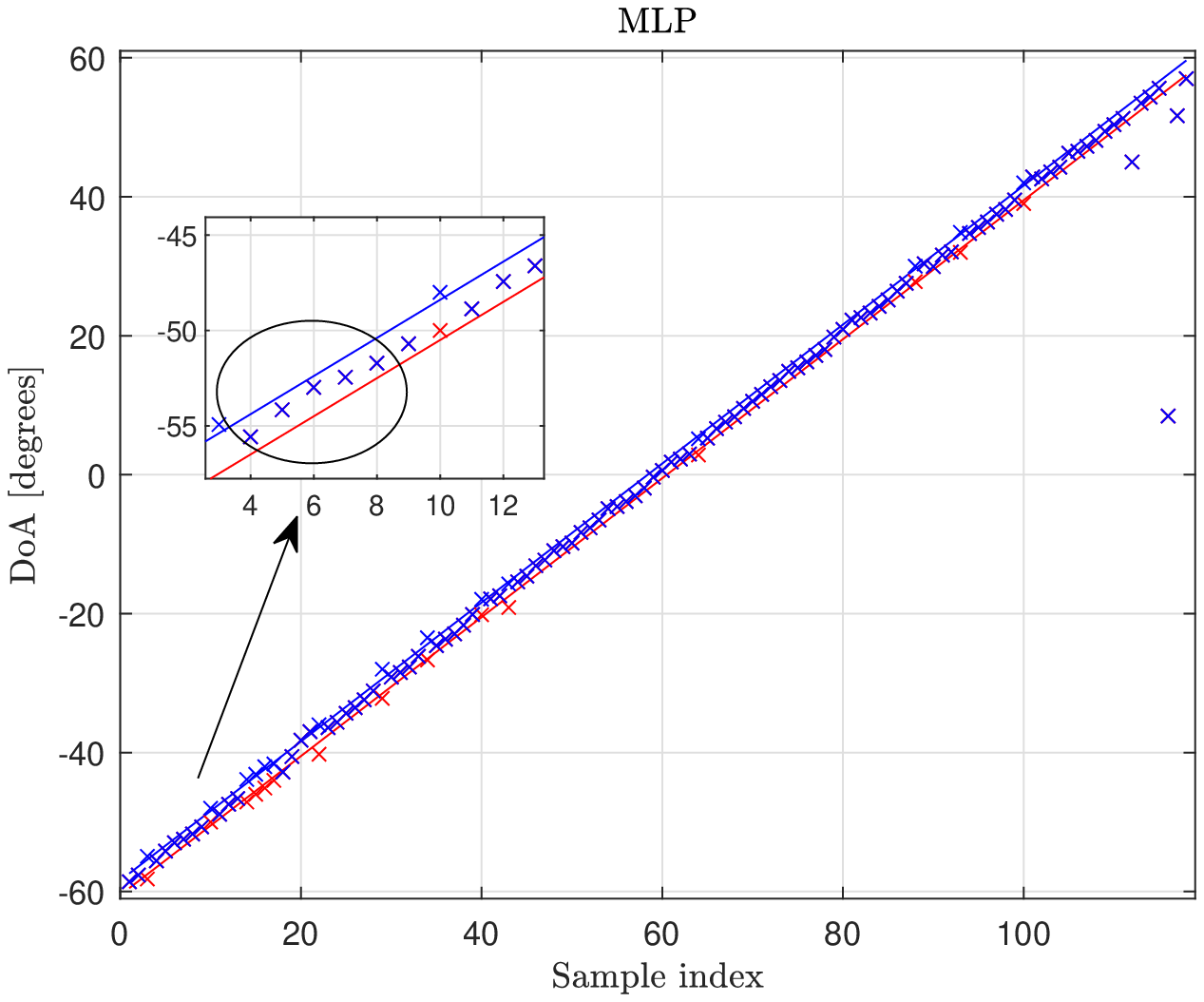}}\hfil   
\subfloat[]{\includegraphics[width=0.25\textwidth]{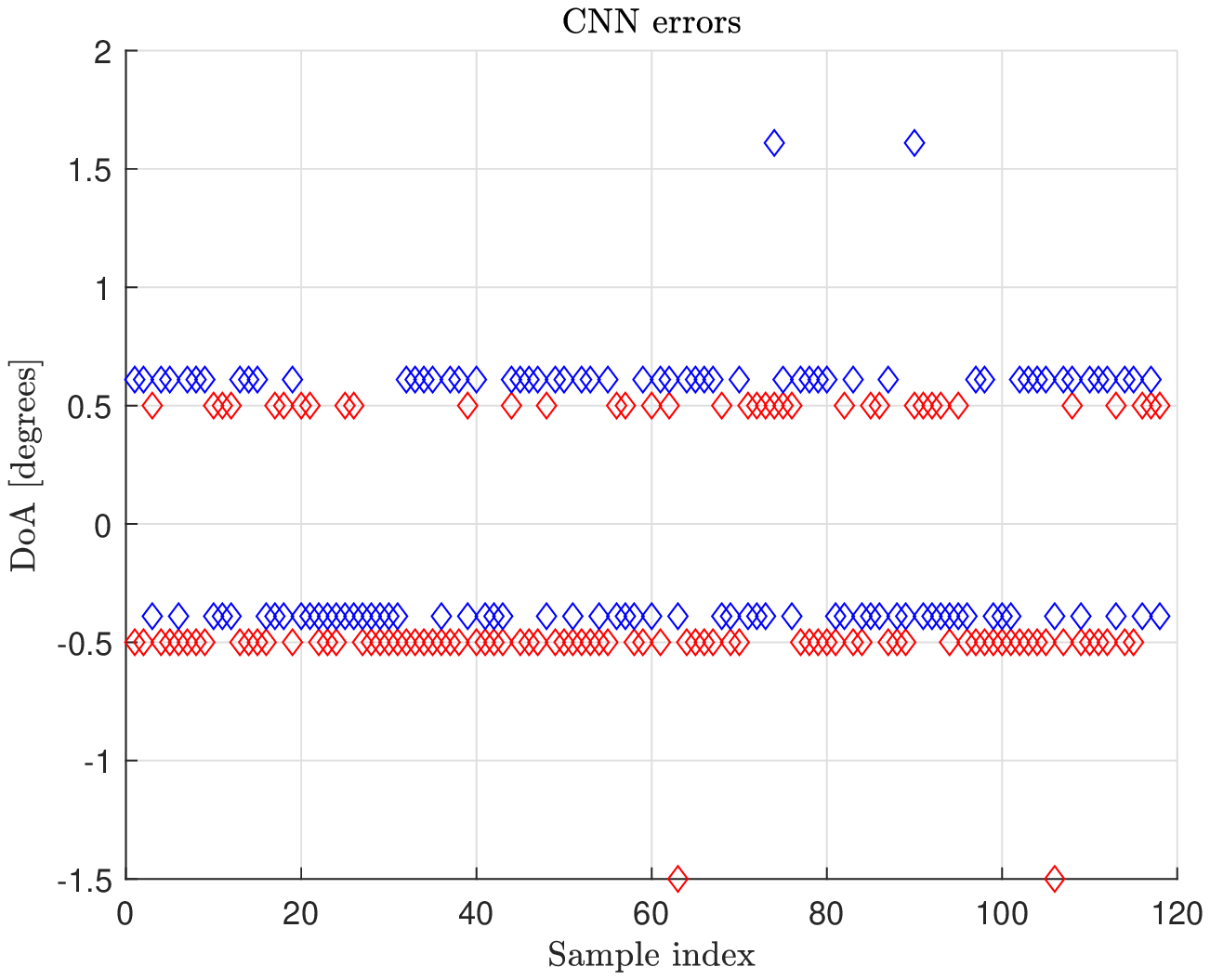}}\hfil
\subfloat[]{\includegraphics[width=0.25\textwidth]{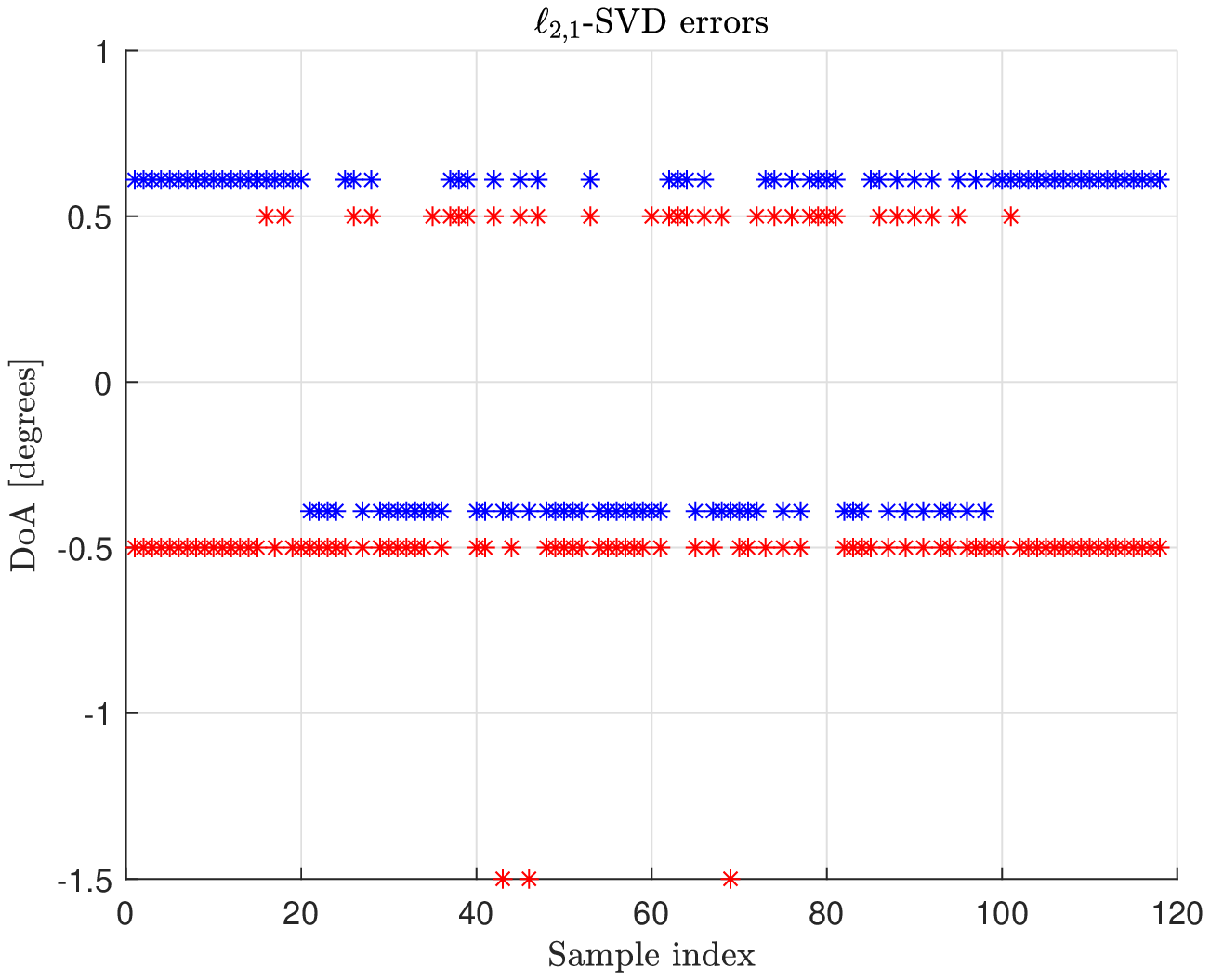}}
\caption{DoA estimation performance on off-grid angles at 0 dB SNR using $T=200$ snapshots. Estimated DoAs by teh (a) CNN (proposed) ,(b) MUSIC, (c) R-MUSIC, (d) $\ell_{2,1}$-SVD and (e) MLP. (f) CNN's errors and (g) $\ell_{2,1}$-SVD's errors. The CNN demonstrates significant gains over the subspace-based methods and similar performance to that of the $\ell_{2,1}$-SVD. More importantly, the additional advantage over the latter method is that no threshold tuning is required.}
\label{fig:Exp1B_0dB_SNR_offgrid_slideangles}
\end{figure*}

Next, we consider two signals with an angular distance of $\Delta \theta = 2.11^\circ$ at $\SNR=0$ dB. The direction of the first signal varies from $-59.5^\circ$ to $57.5^\circ$ (with an increasing step of $1^\circ$). In the second scenario, both signals impinge onto the array from directions unseen during the training procedure of the CNN. The directions are estimated from the sample covariance estimate using $T=200$ snapshots. The CNN's angles estimates are depicted in Fig. \ref{fig:Exp1B_0dB_SNR_offgrid_slideangles}(a) and the corresponding errors are plotted in Fig. \ref{fig:Exp1B_0dB_SNR_offgrid_slideangles}(f). Moreover, in Fig. \ref{fig:Exp1B_0dB_SNR_offgrid_slideangles}(b) and \ref{fig:Exp1B_0dB_SNR_offgrid_slideangles}(c), we have plotted the MUSIC and R-MUSIC DoA estimates, respectively. In Fig. \ref{fig:Exp1B_0dB_SNR_offgrid_slideangles}(d), the $\ell_{2,1}$-SVD DoA estimates are depicted and their corresponding errors are shown in Fig. \ref{fig:Exp1B_0dB_SNR_offgrid_slideangles}(g). Finally, the DoAs estimated  by the MLP are shown in Fig. \ref{fig:Exp1B_0dB_SNR_offgrid_slideangles}(e). Despite the small angular separation between the angles, we observe that the majority of the CNN's errors lie in the interval $[-0.5^\circ,0.61^\circ]$ and only a few errors at $-1.5^\circ$ and $1.61^\circ$. On the other hand, the MUSIC estimator has several large errors and the R-MUSIC has only a few angle errors close to the boundaries of the angular region, i.e., $\{-60^\circ,60^\circ\}$. The performance of the $\ell_{2,1}$-SVD is shown to be more robust than the subspace-based methods. The RMSE in the DoA estimation is $0.54^\circ$ for the CNN and the $\ell_{2,1}$-SVD, $20.31^\circ$ for MUSIC and $11.17^\circ$ for R-MUSIC. Albeit the R-MUSIC demonstrates an overall good performance, a few very large errors occur at the beginning and end of the angular spectrum under study (circled in Fig. \ref{fig:Exp1B_0dB_SNR_offgrid_slideangles}(c)). The MLP demonstrates poor performance and is not able to resolve the closely spaced angles in the current setup (it fails in 103 out of 118 examples). The performance of the proposed CNN is similar to that of the $\ell_{2,1}$-SVD. However, we should note that the tuning of the $\eta$ threshold (noise bound) is one of the major disadvantages of the $\ell_{2,1}$-SVD method, since in many practical applications the operating SNR level is unknown\footnote{There are of course methods that estimate the SNR level, but correspondence to the $\eta$ threshold is not straightforward.}. Hence, we conclude that in contrast to the other approaches, the proposed CNN demonstrates a robust and automated (parameter independent) solution in the DoA estimation of two sources in the low and moderate SNR regime. The threshold value used for the $\ell_{2,1}$-SVD method in this experiment is $\eta=60$.

\subsubsection{RMSE versus the SNR}
\label{sssec:results_rmsevsSNR}

\begin{figure}[t]
\centering
\includegraphics[scale = 0.55]{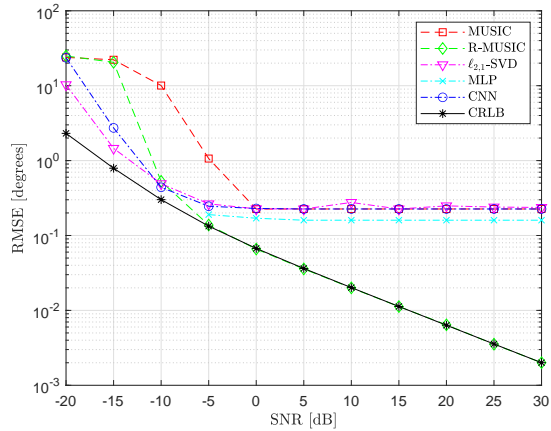}
\caption{The RMSE (logarithmic scale) vs the SNR in the DoA estimation of two sources for off-grid angles using $T=1,000$ snapshots. The proposed CNN outperforms the subspace-based methods in the low-SNR regime. In the high-SNR regime the RMSE of the grid-based approaches floors and only the grid-less R-MUSIC estimator performs optimally and attains the CRLB.}
\label{fig:RMSE_vs_SNR}
\end{figure}

In this experiment, we evaluate the performance of the proposed CNN in the DoA estimation of two sources in the directions $\theta_1=10.11^\circ$ and $\theta_2=13.3^\circ$. At each SNR level the RMSE is calculated over $D_{\text{te}}=1,000$ Monte Carlo (MC) runs, whereas the sample covariance is estimated using $T=1,000$ snapshots. Additionally, we have also calculated the DoA CRLB of the unconditional model introduced in \cite{Stioca1990}. The RMSE results are plotted in Fig. \ref{fig:RMSE_vs_SNR}. In the low-SNR regime, the CNN demonstrates a very well-balanced performance in terms of the RMSE, which is close to that of the robust $\ell_{2,1}$-SVD (except for the lowest SNR value at -20 dB). However, in contrast to the later method, no tuning of any sort of parameters is required for the CNN, which is a major advantage in practical applications if the SNR level is either unknown or it may slightly vary.  The results also indicate that in the high-SNR regime the RMSE floors for the grid-based methods, whereas only the gridless estimator R-MUSIC can attain the CRLB. This performance holds for all grid-based methods and cannot be improved unless a finer grid is utilized. It should be noted that the CNN manages to predict sufficient angle estimates at the high-SNRs despite not being trained in such scenario. Additionally, we have plotted the MLP RMSE results from SNR$=-5$ dB and higher, since at lower SNRs the DoAs could not be consistently resolved. The threshold values used for the $\ell_{2,1}$-SVD method in this experiment are $\eta=\{1260, 700, 400, 230, 140, 100, 70, 70, 60, 60,60\}$ for the corresponding SNR values.

\subsubsection{RMSE versus the number of snapshots $T$}
\label{sssec:results_rmsevsT}

\begin{figure}[t]
\centering
\includegraphics[scale = 0.55]{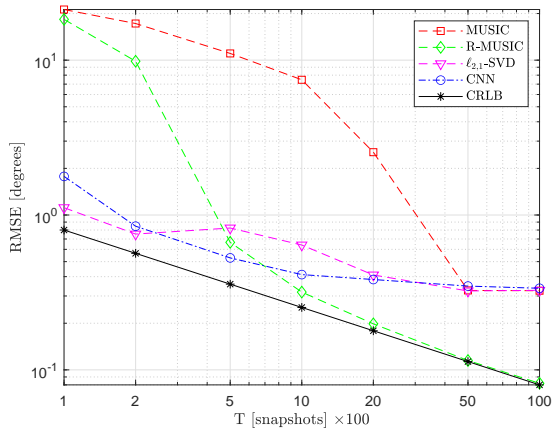}
\caption{The RMSE vs the number of snapshots $T$ (both axes in logarithmic scale) in the DoA estimation of two sources for off-grid angles at $-10$ dB SNR. The proposed CNN demonstrates improved robustness for a small number of snapshots. For a large number of snapshots the performance of grid-based estimators floors, due to the finite grid resolution.}
\label{fig:RMSE_vs_T}
\end{figure}

In this setup, we attempt to estimate the DoAs of two sources at $-10$ dB SNR while the number of snapshots, $T$, varies from $100$ to $10,000$. The direction of the first source is $-13.18^\circ$ and the direction of the second source is $-9.58^\circ$ (off-grid angles). In Fig. \ref{fig:RMSE_vs_T}, the RMSE of the estimation is depicted for each method (both axes are in logarithmic scale); additionally, we have also calculated the CRLB of the DoA estimation. It is observed that for a (relatively) small number of snapshots (up to $T=500$) the CNN demonstrates a robust behavior compared to the subspace based methods MUSIC and R-MUSIC (close to the CRLB). As the number of snapshots increases, the finite resolution of the grid limits the performance of the grid-based estimators. The grid-less R-MUSIC estimator is the only method to perform optimally. Notable is also the fact that the performance of the CNN is similar to that of the $\ell_{2,1}$-SVD method, but with the additional advantage of not depending on the fine-tuning of any parameters that control the estimation. The MLP was not able to resolve the angles in this setup. The threshold values used for the $\ell_{2,1}$-SVD method in this experiment are $\eta=\{130, 180, 270, 410, 570, 910, 1280\}$ for the corresponding values of $T$.

\subsubsection{RMSE versus the angle separation $\Delta \theta$}
\label{sssec:results_rmsevsangsep}

\begin{figure}
\centering
\includegraphics[scale = 0.55]{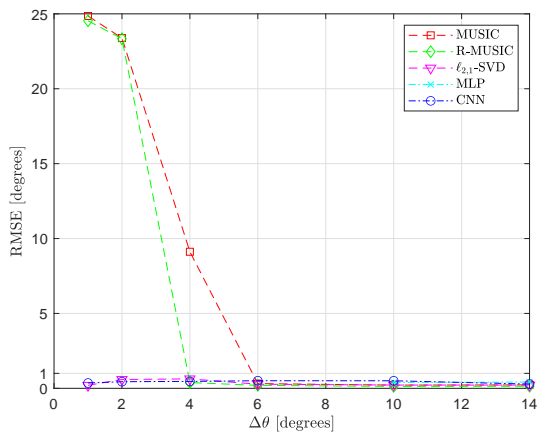}
\caption{The RMSE vs the angle separation $\Delta \theta$ in the DoA estimation of two sources for off-grid angles at $-10$ dB SNR using $T=500$ snapshots. The small angular separation of the sources does not affect the performance of the CNN and the $\ell_{2,1}$-SVD.}
\label{fig:RMSE_vs_ang_sep}
\end{figure}

Fig. \ref{fig:RMSE_vs_ang_sep} includes RMSE results in the DoA estimation of two sources for various angle separations $\Delta \theta \geq 1^\circ$. The DoA of the first source is $\theta_1=-13.8^\circ$ (off-grid) and the second direction is $\theta_2=\theta_1+\Delta \theta,$ while SNR$=-10$ dB and the number of snapshots used is $T=500$. We observe that for closely separated angles ($1^\circ \leq \Delta \theta <4^\circ$) the CNN and the $\ell_{2,1}$-SVD are able to resolve the angles, as opposed to the subspace-based methods, which fail to provide accurate DoA estimates. Moreover, the performance of the CNN and the $\ell_{2,1}$-SVD is almost constant for all angular separations, with an RMSE less than $0.65^\circ$. The MLP can only resolve DoAs with angle separation $\Delta \theta>6^\circ$ (therefore only results for $10^\circ$ and $14^\circ$ are provided). The threshold value used for the $\ell_{2,1}$-SVD in this experiment is $\eta=290.$

\subsubsection{Robustness to SNR Mismatches}
\label{sssec:robst_snr_mism}

In all previous simulations, at each SNR we set $\sigma_1^2=\sigma_2^2=1$ and calculated $\sigma_e^2$ according to Eq. \eqref{eq:SNR}. However, perfect knowledge of the exact SNR level is not always guaranteed in practice. In this setup, we evaluate the robustness of the proposed method against SNR mismatches by comparing the CNN's errors to those of the $\ell_{2,1}$-SVD. Two scenarios are considered here as well. 

\begin{figure*}
\centering
\subfloat[]{\includegraphics[width=0.32\textwidth]{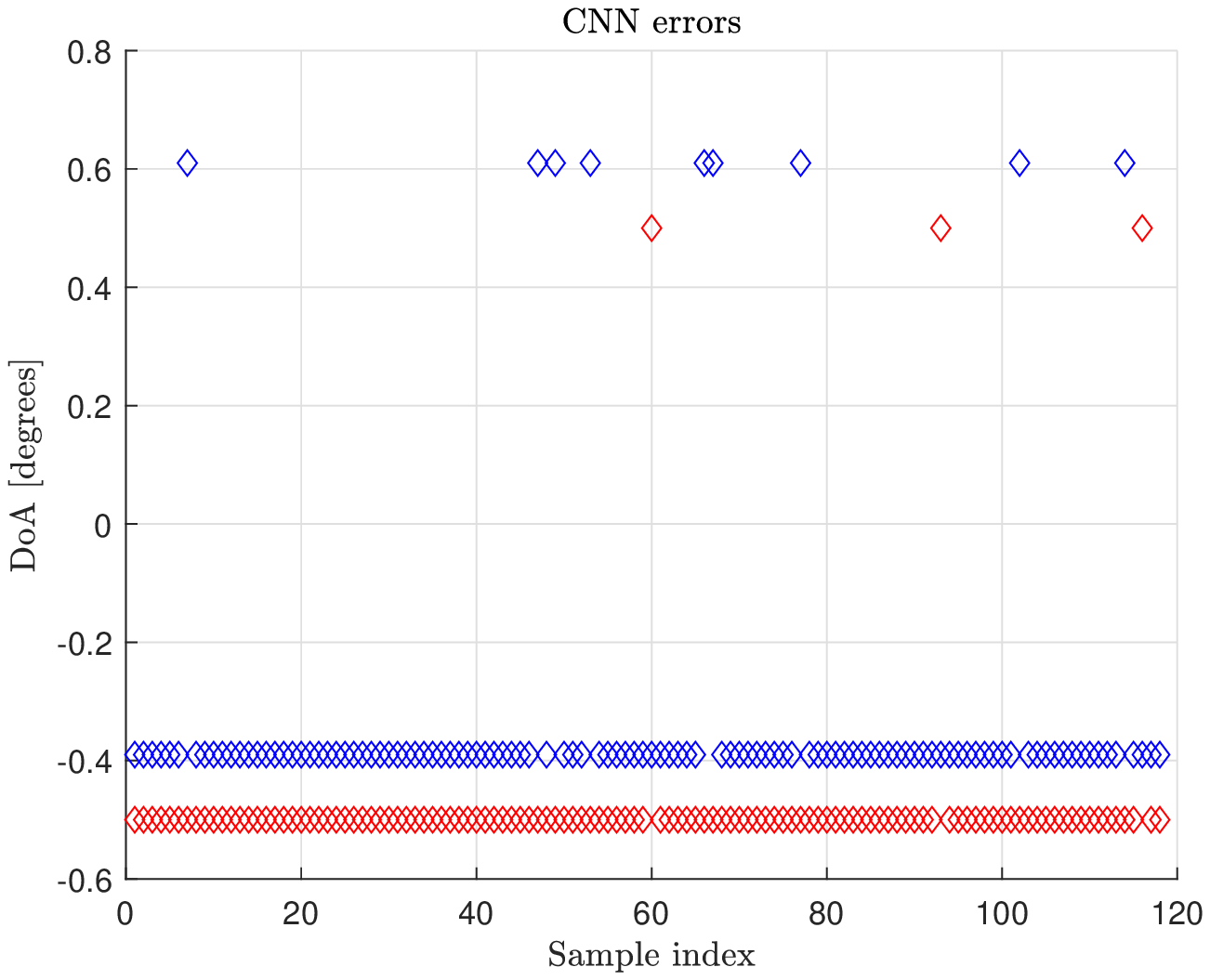}}\hfil
\subfloat[]{\includegraphics[width=0.32\textwidth]{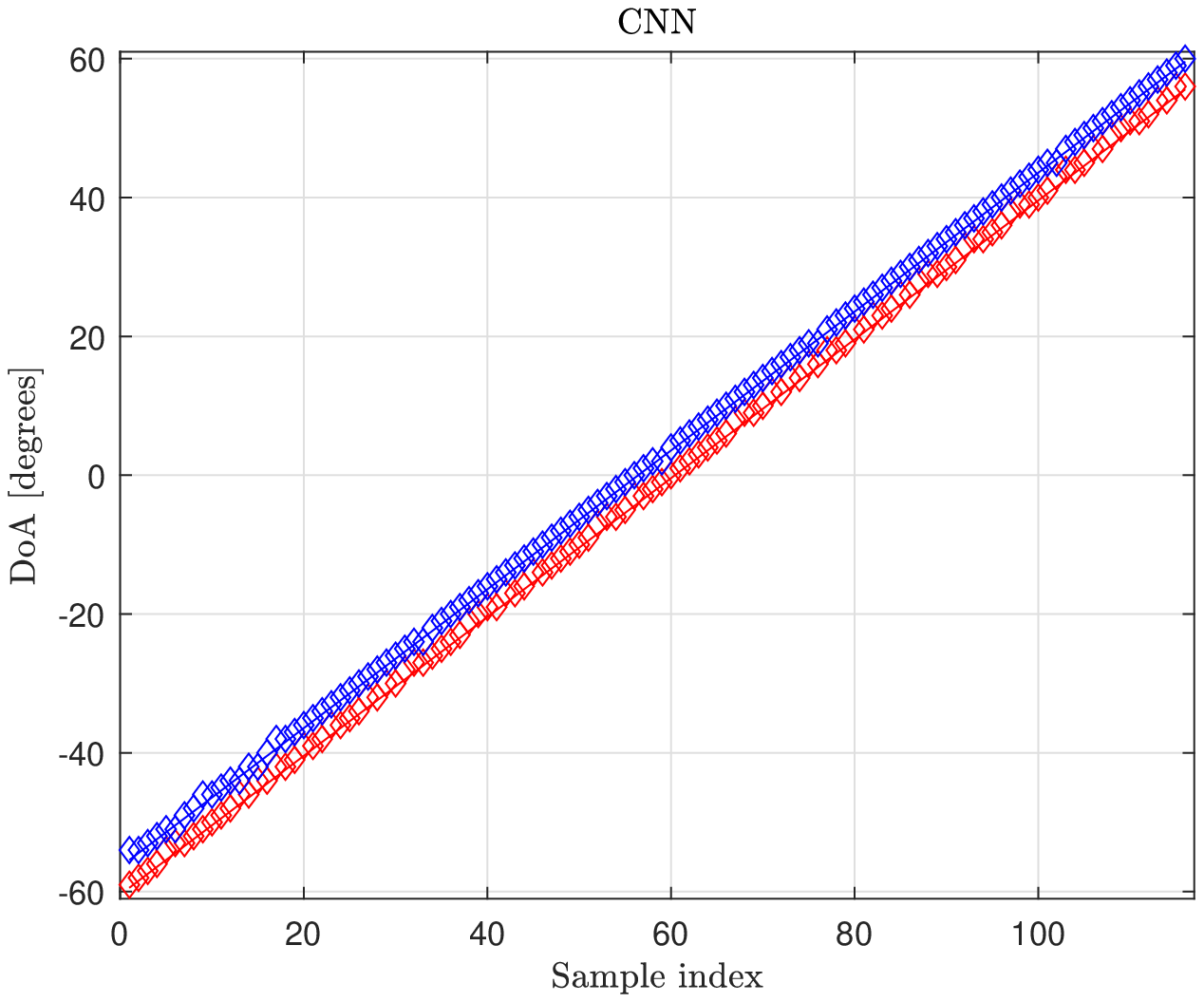}}\hfil 
\subfloat[]{\includegraphics[width=0.32\textwidth]{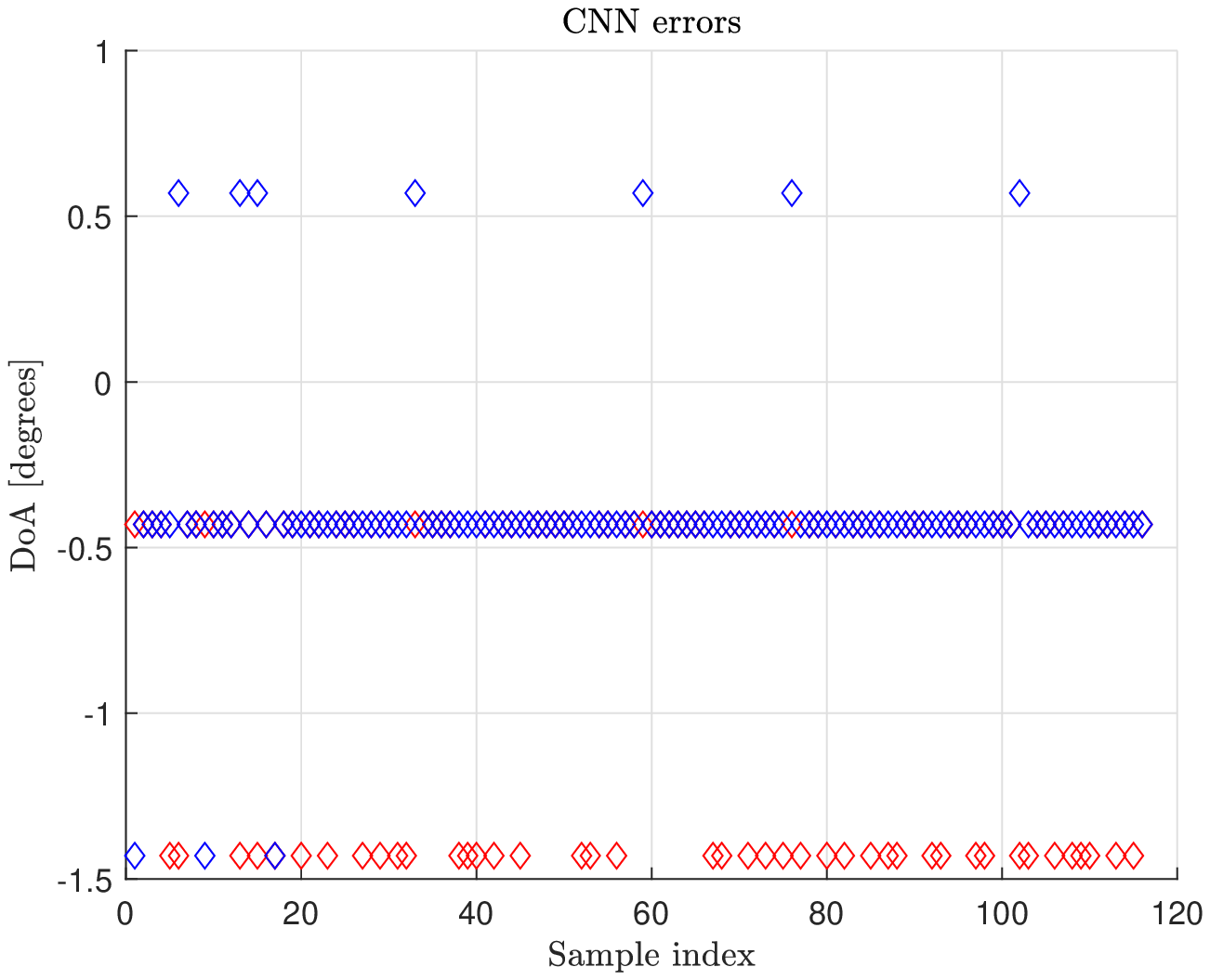}} 

\subfloat[]{\includegraphics[width=0.32\textwidth]{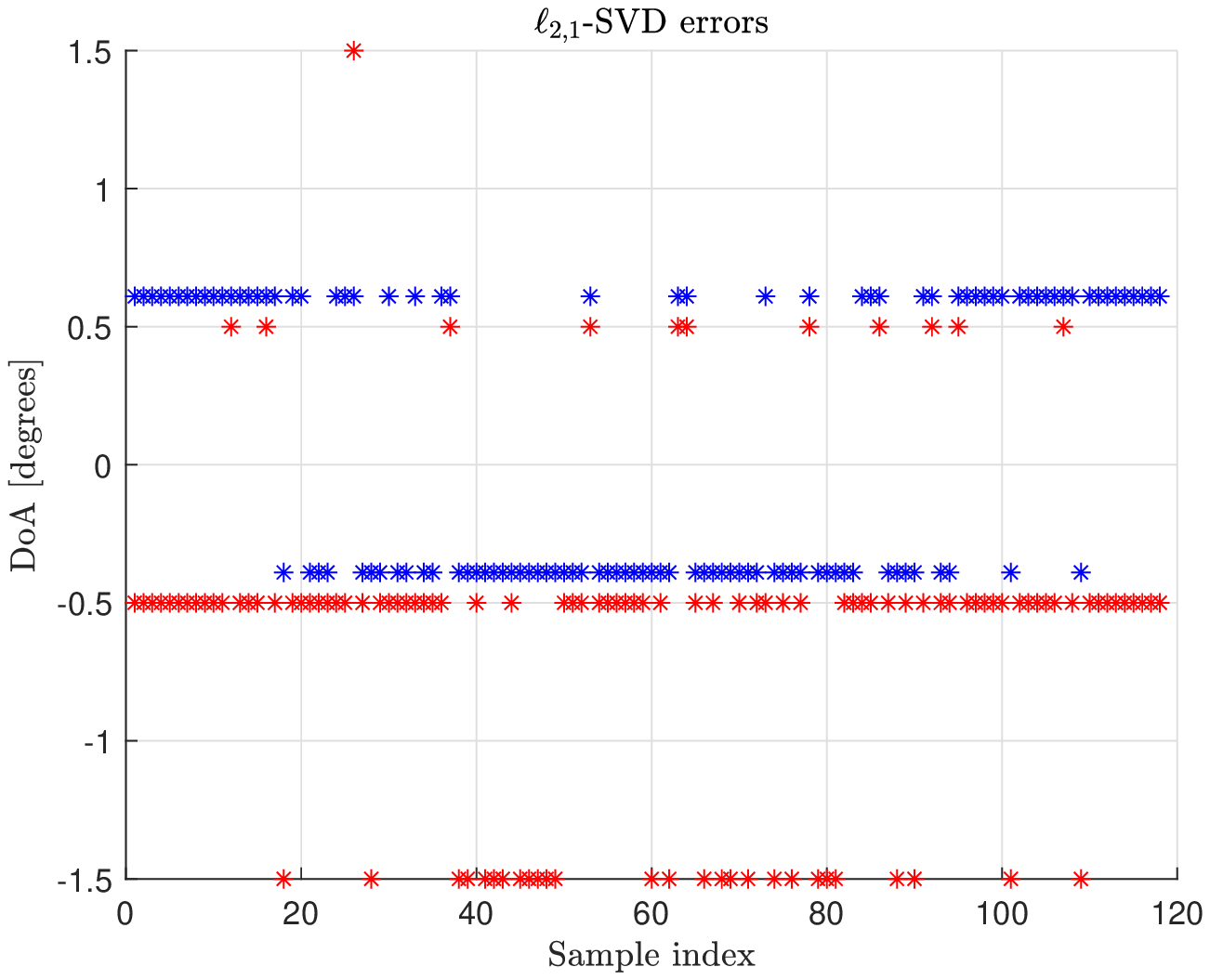}}\hfil   
\subfloat[]{\includegraphics[width=0.32\textwidth]{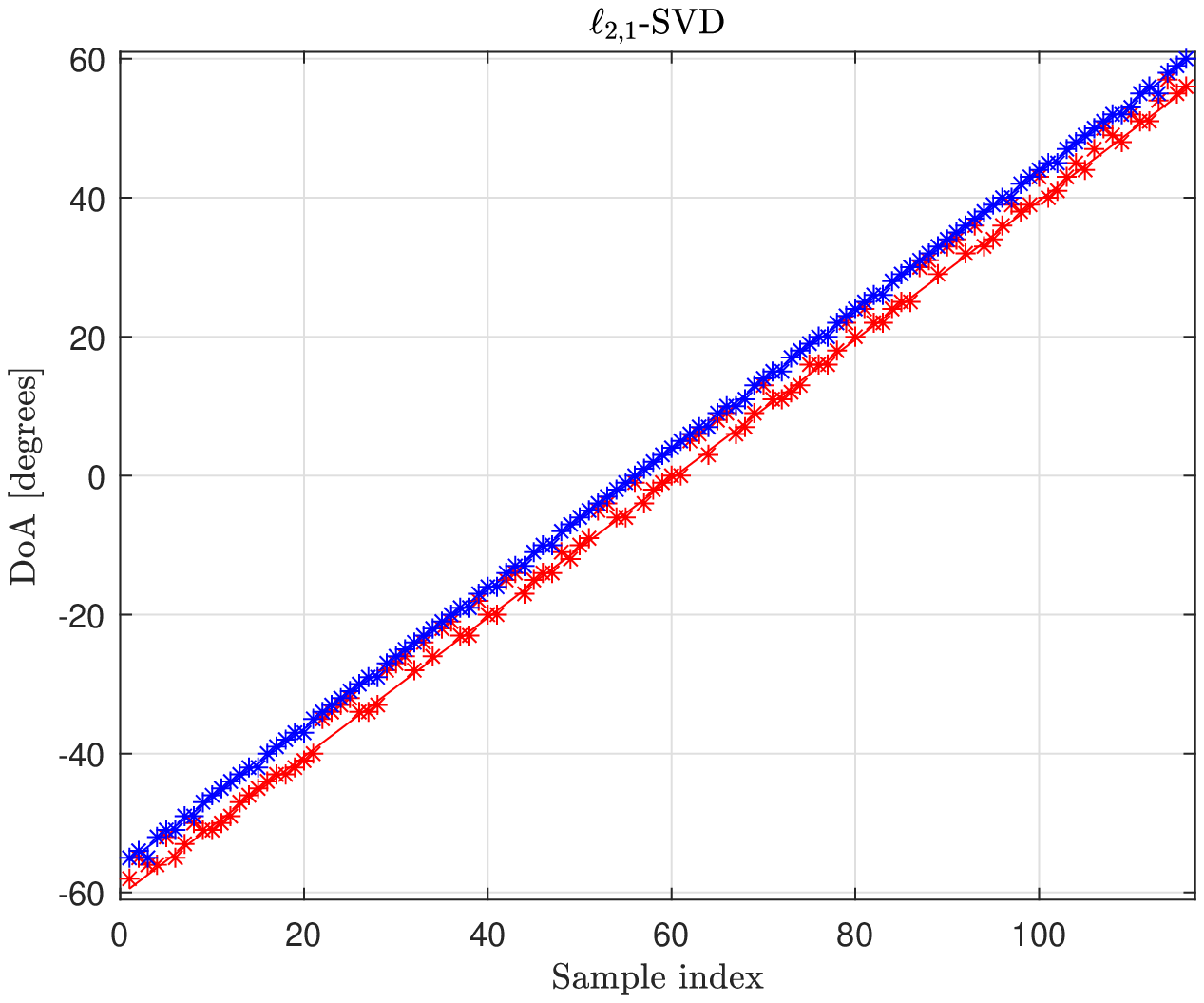}}\hfil
\subfloat[]{\includegraphics[width=0.32\textwidth]{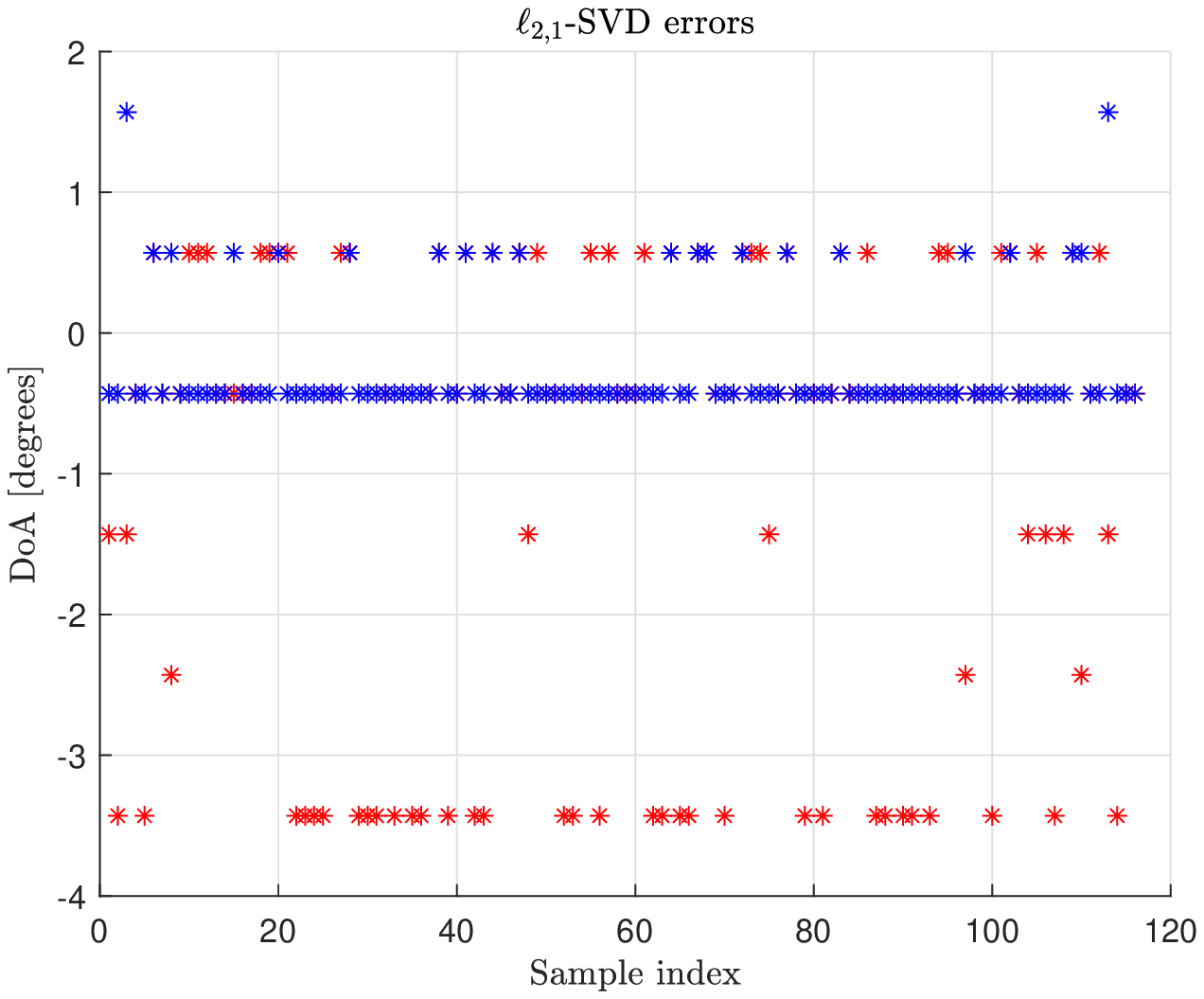}}
\caption{Comparison between the CNN (proposed) and the $\ell_{2,1}$-SVD methods in the DoA estimation of two sources under SNR mismatches. CNN results in (a-c) and $\ell_{2,1}$-SVD results in (d-f). (a), (d) Errors at SNR$=0$ dB using $T=200$ snapshots. (b), (e) DoA estimates at SNR$=-10$ dB using $T=1,000$ snapshots and the respective errors in (c), (f).}
\label{fig:Exp5}
\end{figure*}

The first one closely follows the setup of the second experiment in Section \ref{sssec:error_distr} at 0 dB SNR (the same angular directions and number of snapshots $T=200$). The noise variance is $\sigma_e^2=1$; however, instead of unit power, we have considered a small perturbation to the sources' power with  $\sigma_1^2=0.7$ and $\sigma_2^2=1.25$. As a result, the actual SNR is now $-1.549$ dB instead of $0$ dB. The errors in the DoA estimation of the two sources are depicted in Fig. \ref{fig:Exp5}(a) for the CNN and in Fig. \ref{fig:Exp5}(d) for the $\ell_{2,1}$-SVD. For the latter method the threshold value $\eta=60$ was optimized for use at $0$ dB SNR. We observe a larger dispersion of errors for the $\ell_{2,1}$-SVD, whereas the CNN demonstrates enhanced robustness. The RMSE of the estimation is $0.46^\circ$ and $0.7^\circ$ for the CNN and the $\ell_{2,1}$-SVD, respectively. 

In the second experiment, we consider two angular directions separated by $4^\circ$. The direction of the first source ranges from $-59.43^\circ$ to  $55.57^\circ$ with an increasing step of $1^\circ$, whereas the number of snapshots is $T=1,000$. We consider $\sigma_e^2=10$ and produce a small perturbation to the sources' power with  $\sigma_1^2=0.7$ and $\sigma_2^2=1.25$, which leads to an actual SNR$=-11.549$ dB instead of $-10$ dB. The DoA estimates and the errors of the CNN are depicted in  Fig. \ref{fig:Exp5}(b) and \ref{fig:Exp5}(c), respectively, whereas the estimates and errors of the $\ell_{2,1}$-SVD method are depicted in  Fig. \ref{fig:Exp5}(e) and \ref{fig:Exp5}(f), respectively. The empirical RMSE of the estimation is $1.42^\circ$ for the $\ell_{2,1}$-SVD and $0.74^\circ$ for the CNN, leading to an improvement of $48\%$. The noise bound for the CS-based method was set to $\eta=400$ (optimized for use at $-10$ dB SNR). We conclude that the sensitivity of the $\ell_{2,1}$-SVD method to SNR mismatches is high and this is one of the additional advantages of the proposed CNN.

\subsection{Number of Sources Unknown}
\label{ssec:unknownK}

In the second part of the section, after training the network according to Section \ref{ssec:scenario2}, we test the CNN without using information about the number of transmitting sources. Thus, we relax the assumption that the number of sources is known a priori and only consider that their maximum number, $K_{max}=3$, is known (which is only required for the training). After training the model, we can use a confidence level $\bar{p}$ and select the grid points $i=1,\dots, 2G+1$ with probability $\hat{p}_i \geq \bar{p}$ in Eq. \eqref{eq:predictor}. However, there is no guarantee that the number of source DoAs will now be equal to $K$; as a matter of fact, the cardinality of the predicted sets of directions may vary from zero to $\hat{K}_{max} $ (it may also hold that $\hat{K}_{max}\lesseqqgtr K_{max}$). Hence, the RMSE metric is no longer suitable for the evaluation of the CNN's performance. To this end, we resort to the Hausdorff distance, which for two sets $\mathcal{A}$ and $\mathcal{B}$ is defined as:
\begin{equation}
d_{\text{H}}(\mathcal{A}, \mathcal{B}) = \max \{ d(\mathcal{A}, \mathcal{B}), d(\mathcal{B}, \mathcal{A}) \},
\label{eq:Hausdorff_distance_sets}
\end{equation}
where 
\begin{equation}
d(\mathcal{A}, \mathcal{B}) = \sup \{ d(\alpha, \mathcal{B})| \alpha \in \mathcal{A}\}
\label{eq:d_directed_distance}
\end{equation}
is the directed difference\footnote{Notice that in general $d(\mathcal{A}, \mathcal{B}) \neq d(\mathcal{B}, \mathcal{A})$.}, $d(\alpha, \mathcal{B})= \inf \{d(\alpha, \beta)|\beta \in \mathcal{B}\}$ and $d(\alpha,\beta) = |\alpha-\beta|.$ For the evaluation over the testing set we have used its mean and maximum value, denoted as $\mu (d_{\text{H}})$ and $\max (d_{\text{H}})$, respectively. Simply stated, the Hausdorff distance measures how far two subsets of a metric space are from each other and is particularly important in cases where $\mathcal{A}$ and $\mathcal{B}$ do not share equal cardinality. For example, if $\mathcal{A}=\{-30^\circ, 20^\circ, 23^\circ \}$ and $\mathcal{B}=\{-30.2^\circ, 20.15^\circ, 22.83^\circ \}$ their $\RMSE = 0.18^\circ$ whereas $d_{\text{H}}(\mathcal{A}, \mathcal{B})=0.2^\circ$ (which is the maximum error corresponding to the first angle). Moreover, if $\mathcal{A}=\{-30^\circ, 21^\circ \}$ the Hausdorff distance is then $d_{\text{H}}(\mathcal{A}, \mathcal{B})=1.83^\circ$. On the other hand, if $\mathcal{A}=\{-30^\circ, 51^\circ \}$ the Hausdorff distance becomes $d_{\text{H}}(\mathcal{A}, \mathcal{B})=30.85^\circ$. The latter example shows that large deviations are severely penalized by the metric.

\begin{figure}
\centering
\subfloat[SNR$=-10$ dB]{\includegraphics[width=0.4\textwidth]{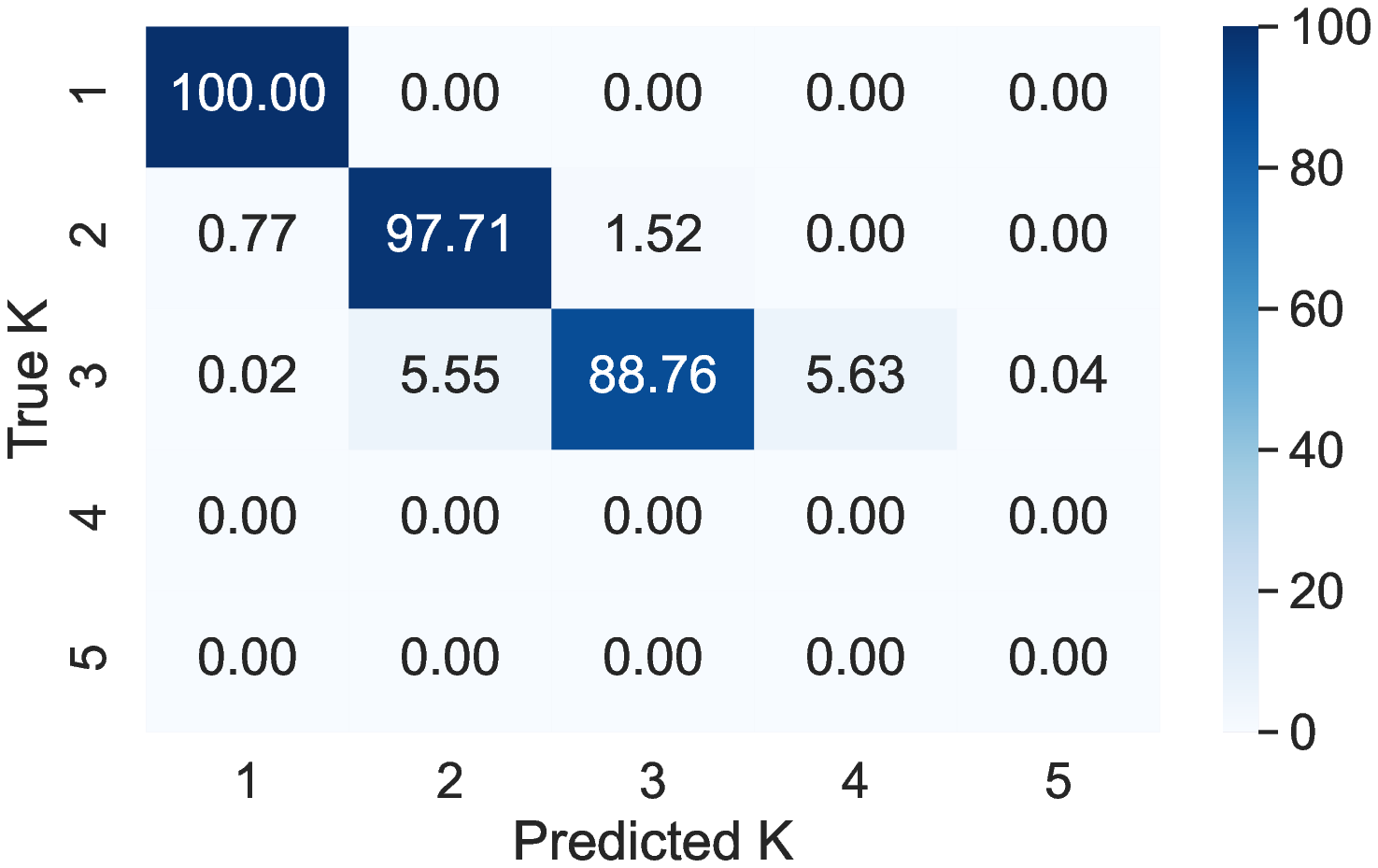}}\hfil
\subfloat[SNR$=0$ dB]{\includegraphics[width=0.4\textwidth]{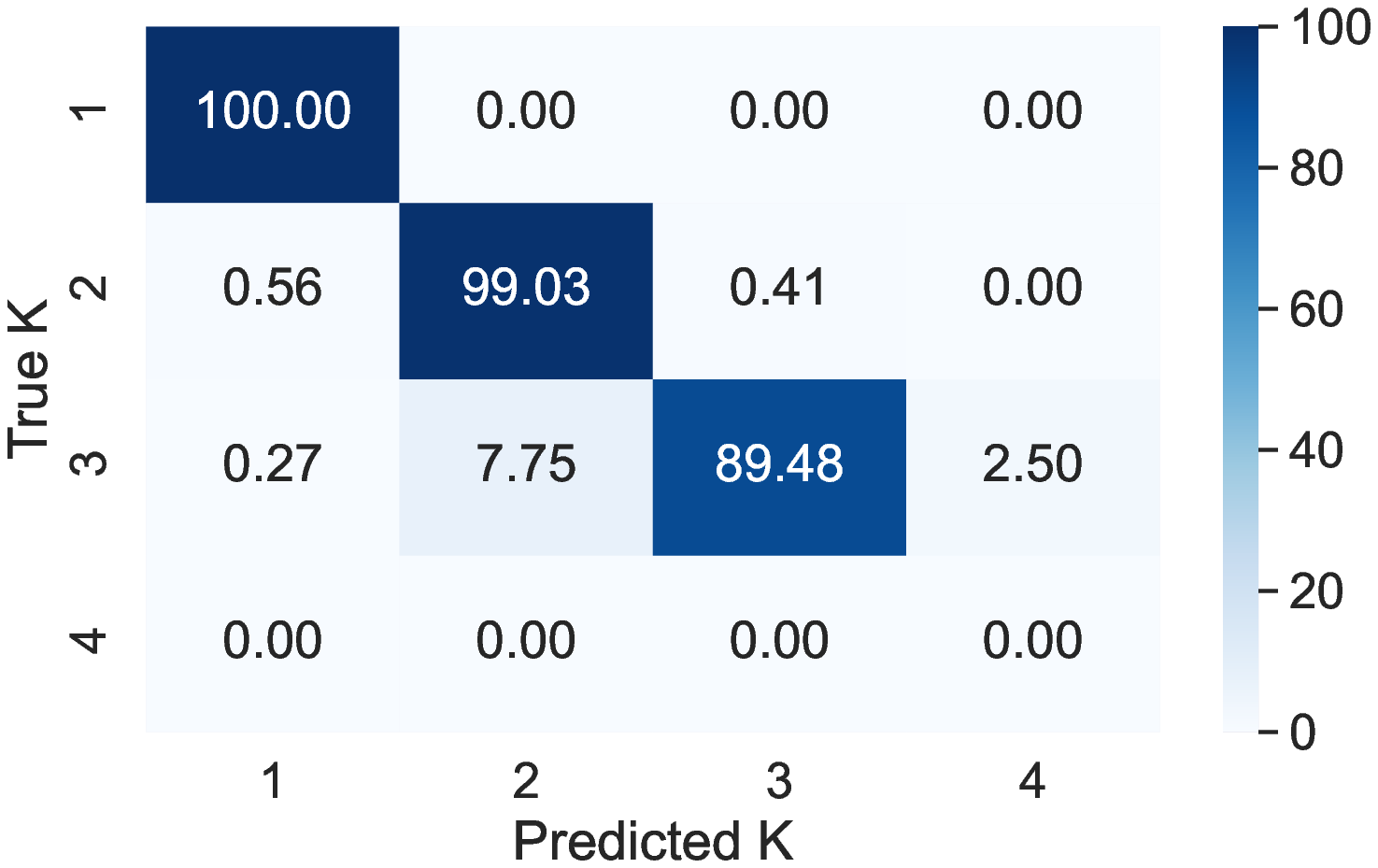}}\hfil
\caption{Confusion matrix results (\%) on the number of sources classification using (a) $T=3,000$ and (b) $T=1,000$ snapshots for the sample covariance estimate. At both SNRs there is a very low error rate for up to $K=2$, which increases with the number of sources.}
\label{fig:confusion_matrix}
\end{figure}

First, we consider fixed off-grid angles for $K=1$ up to $K=3$. The direction of the first signal is $7.8^\circ$; the second signal's DoA is at $-2.6^\circ$, and the direction of the third signal is at $2.6^\circ$ (the angle separation is $\Delta \theta=5.2^\circ$). We evaluate the performance of the CNN in terms of the Hausdorff distance with a) $\mu(d_{\text{H}})$, b) $\max (d_{\text{H}})$, as well as c) in the classification of the number of sources, which is unknown. For each $K=1,2,3$ we generated 10,000 testing examples using $T=3,000$ and $T=1,000$ snapshots at $-10$ and $0$ dB SNR, respectively. The results of the DoA estimation are reported on Tab. \ref{tab:Hausdorff_dis_vs_RMSE} for each SNR level. On the second column we provide the selected confidence level $\bar{p}$; on the third column we have listed the mean Hausdorff distance denoted as $\mu (d_{\text{H}})$; the maximum Hausdorff distance denoted as $\max (d_{\text{H}})$ is listed on the fourth column of the table; finally, on the last column, we have included (for comparison) the RMSE of the estimation assuming that the number of transmitting sources $K$ is known in each case. Notice the proximity of the results between the RMSE and $\mu(d_{\text{H}}).$ Additionally, in Fig. \ref{fig:confusion_matrix} we have evaluated the network's ability to classify the number of sources via the confusion matrix, for each SNR level, (a) at -10 dB and (b) at 0 dB (\%) with the confidence levels listed in Tab. \ref{tab:Hausdorff_dis_vs_RMSE}. Entries on the main diagonal correspond to correct/true predictions. Nonzero entries on the right of the main diagonal correspond to Type I or false positive (FP) errors, whereas entries on the left of the main diagonal correspond to Type II or false negative (FN) errors. We observe that, in the majority of the examples, the number of transmitting sources has been correctly identified by the proposed CNN for each $K=1,2,3$. However, we observe that the errors increase with the number of sources, something which is well expected, since the problem is considerably more difficult to solve, especially in the low-SNRs.

\begin{figure*}
\centering
\subfloat[]{\includegraphics[width=0.32\textwidth]{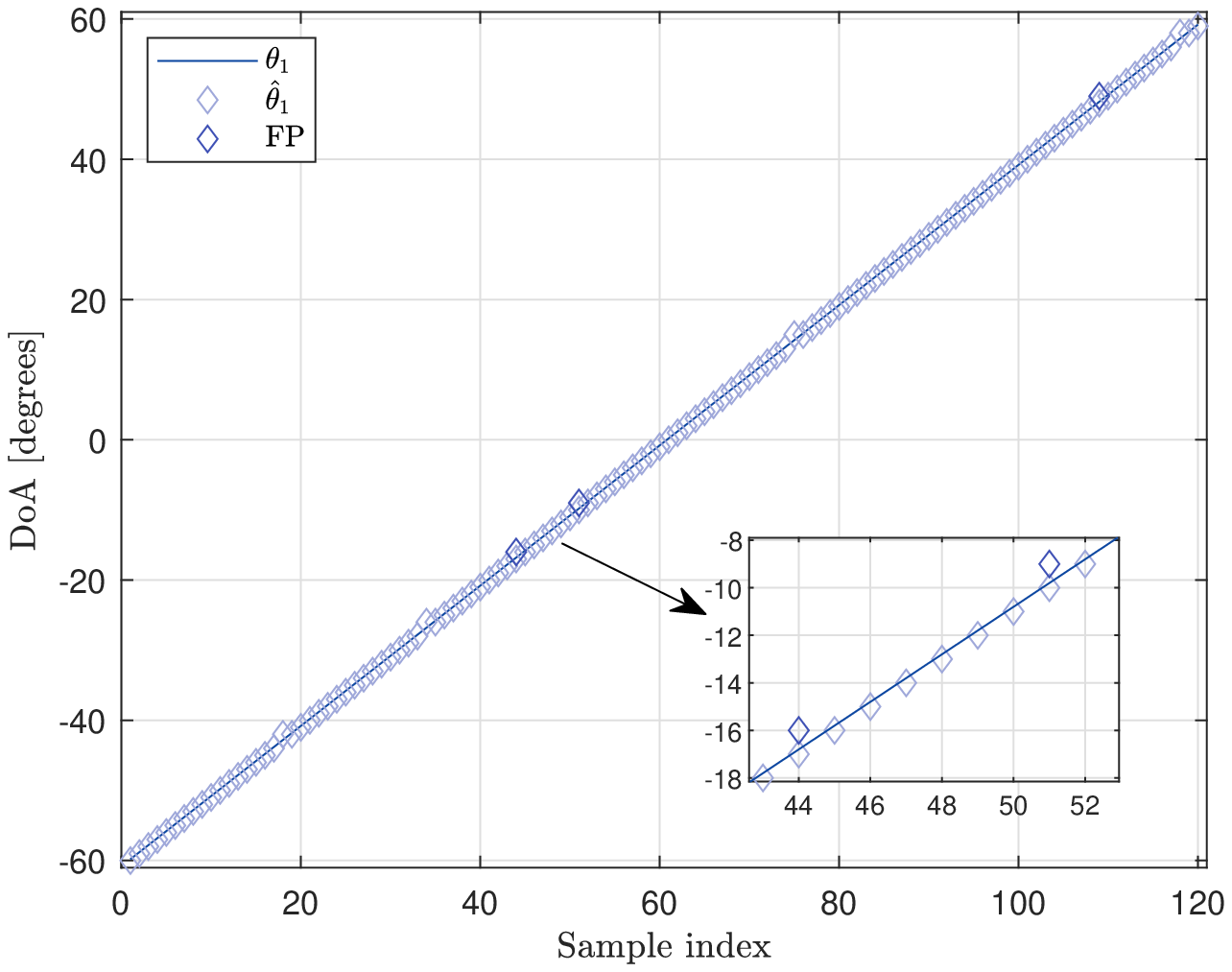}}\hfil
\subfloat[]{\includegraphics[width=0.32\textwidth]{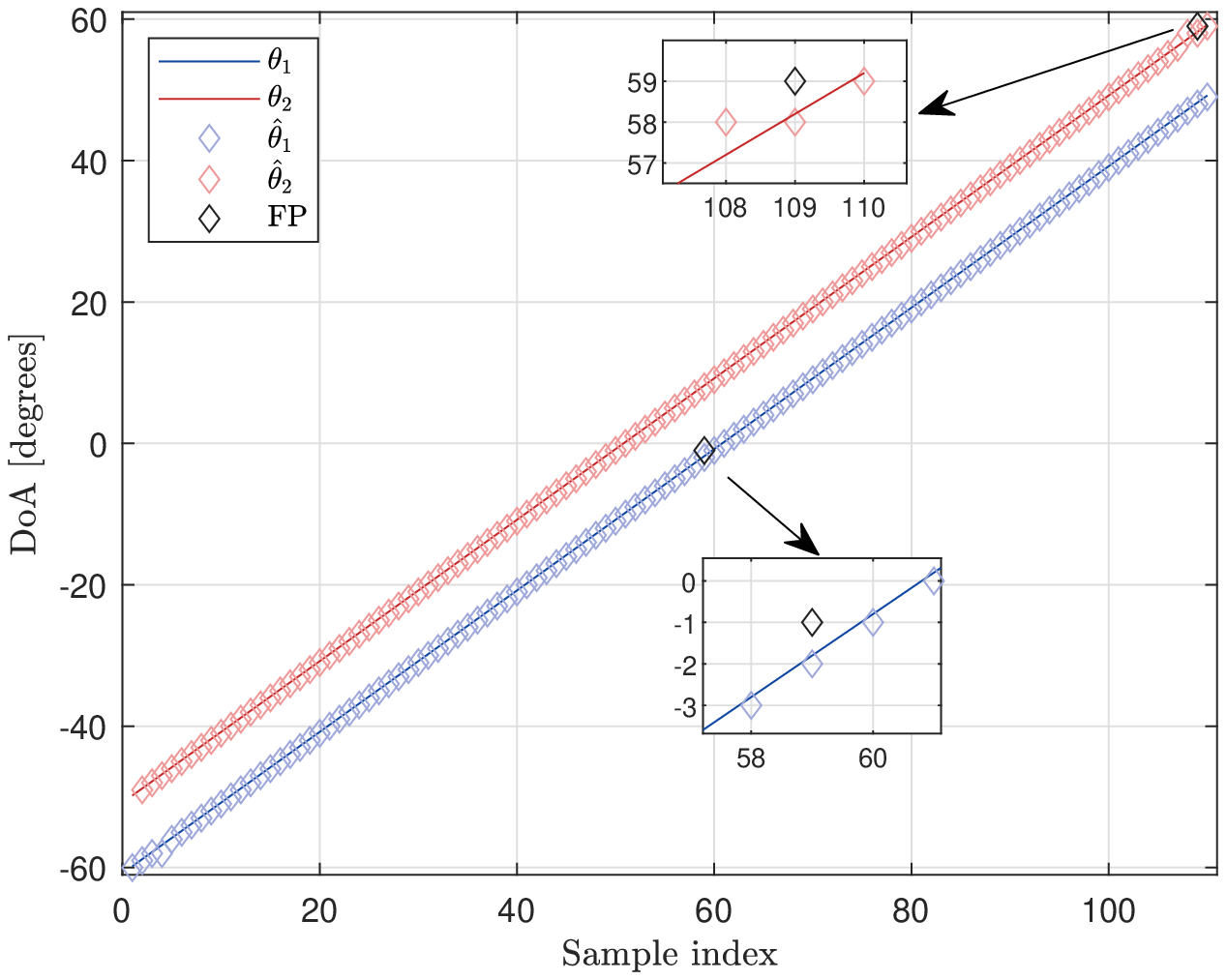}}\hfil 
\subfloat[]{\includegraphics[width=0.32\textwidth]{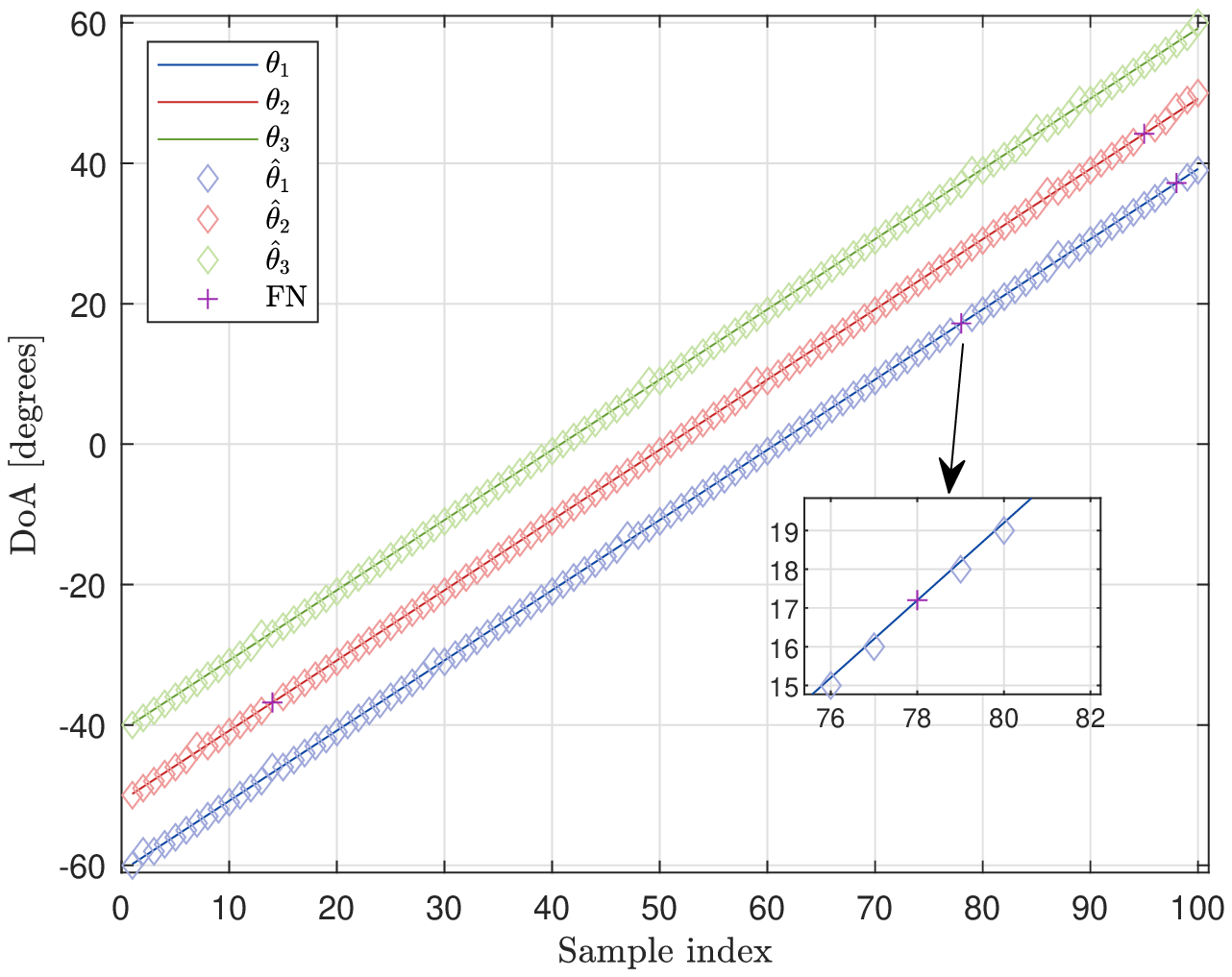}} 

\subfloat[]{\includegraphics[width=0.32\textwidth]{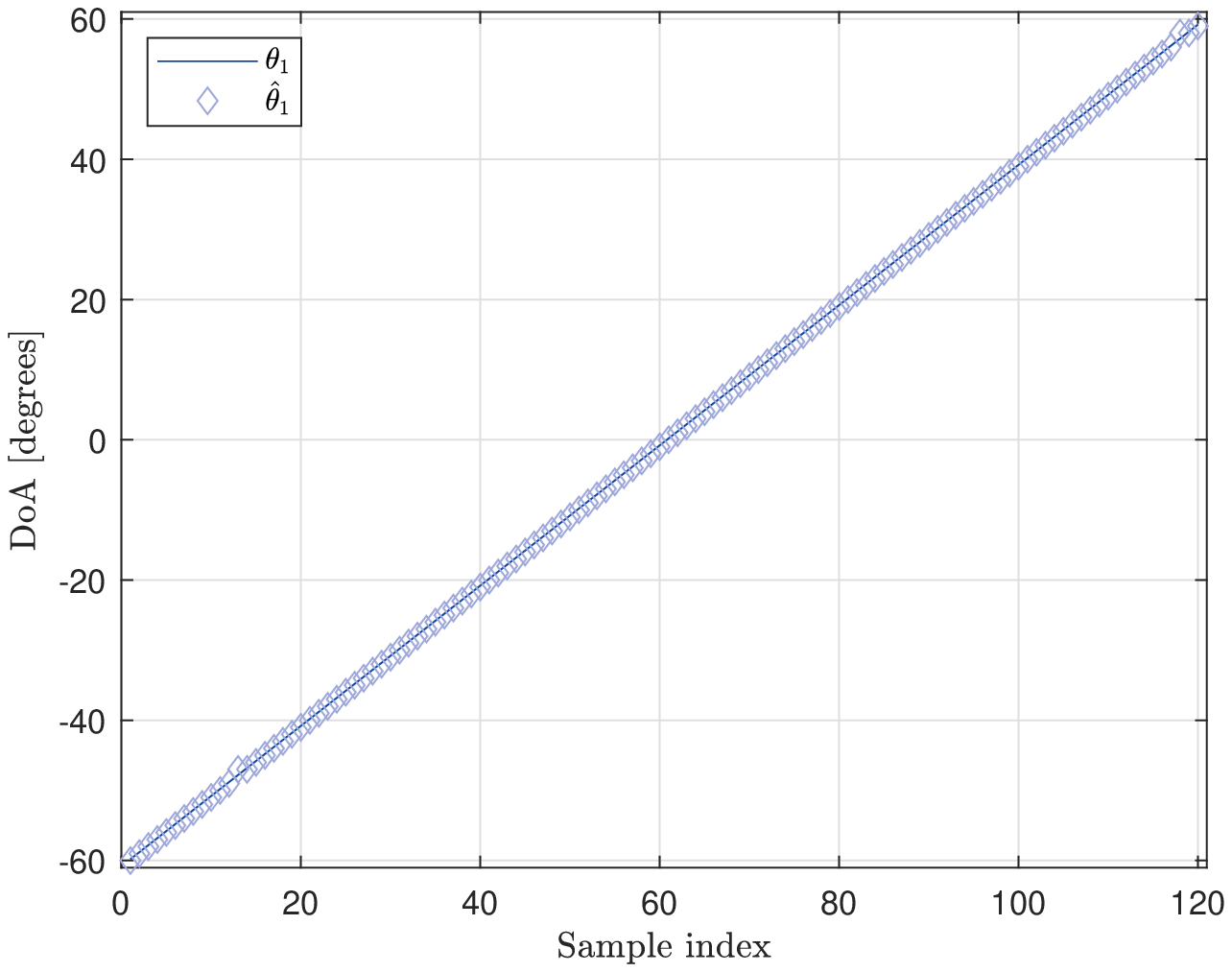}}\hfil   
\subfloat[]{\includegraphics[width=0.32\textwidth]{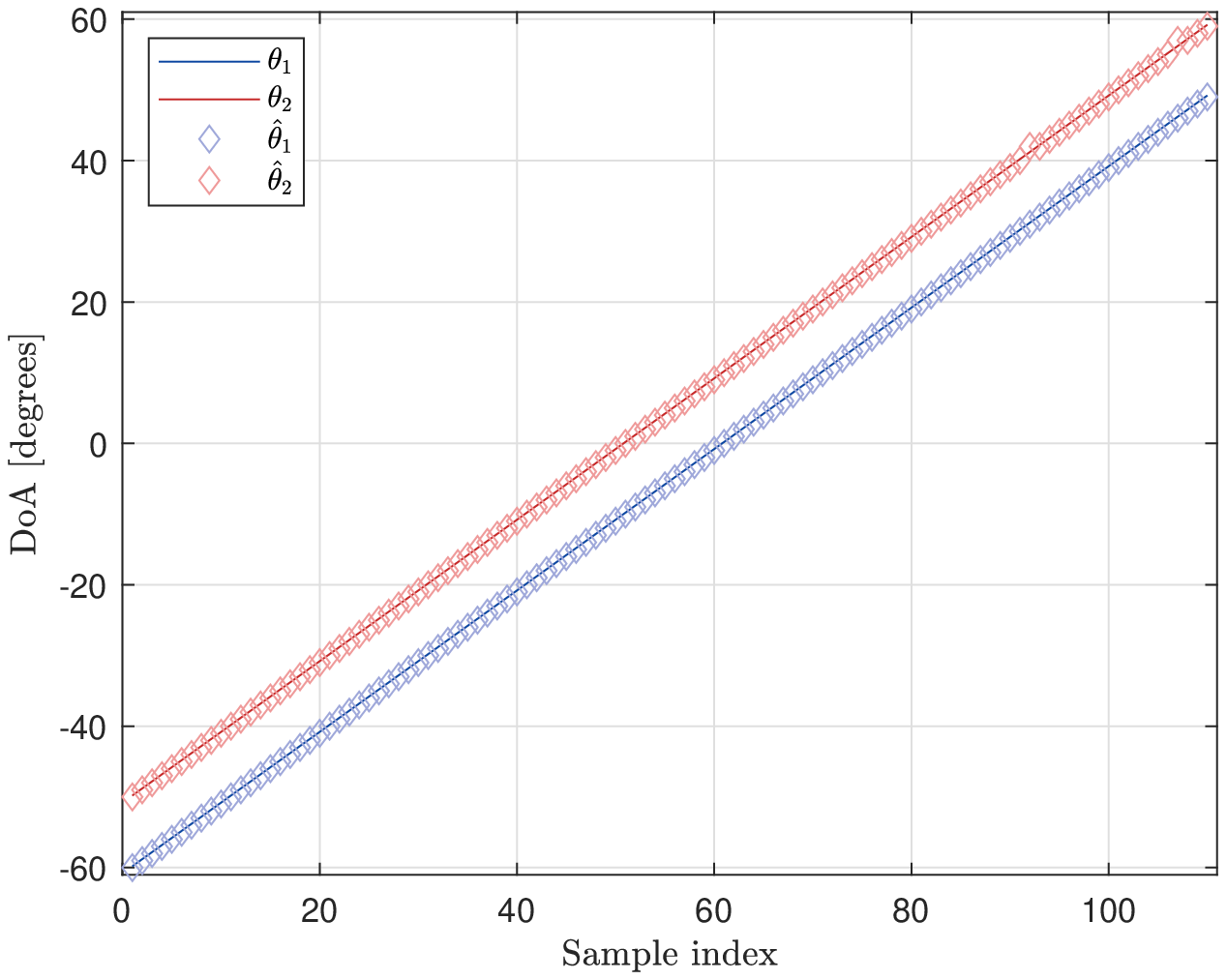}}\hfil
\subfloat[]{\includegraphics[width=0.32\textwidth]{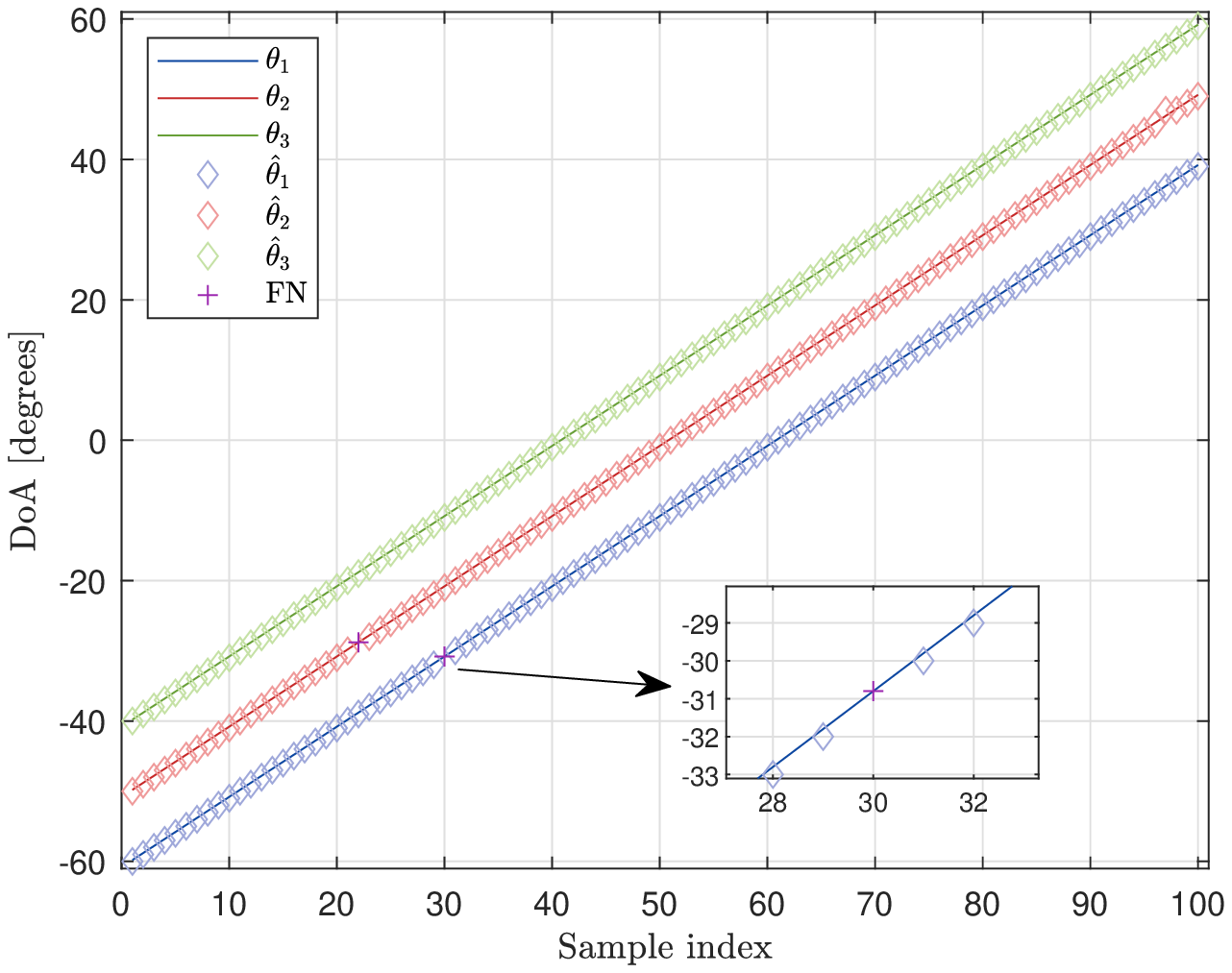}}
\caption{DoA prediction by the CNN without a priori knowledge on the number $K$ of transmitting sources for varying angles. (a-c) corresponds to SNR$=-10$ dB using $T=3,000$ snapshots. (d-f) corresponds to SNR$=0$ dB with $T=1,000$ snapshots (two different networks trained at each SNR). In (a), (d) $K=1$; in (b), (e) $K=2$ and in (c), (f) $K=3$.}
\label{fig:Exp6B}
\end{figure*}

In the final experiment, we let the directions of the sources vary across the angular region of interest. Specifically, for $K=1$ we consider 120 examples of the signal with directions from $-59.8^\circ$ to $59.2^\circ$ and an increasing step of $1^\circ$. For $K=2$, the first signal's direction ranges from $-59.8^\circ$ to $49.2^\circ$ and the second one's from $-49.8^\circ$ to $59.2^\circ$ (110 examples with step $1^\circ$). For $K=3$, the first DoA ranges from $-59.8^\circ$ to $39.2^\circ$, the second one from $-49.8^\circ$ to $49.2^\circ$ and the third one from $-39.8^\circ$ to $59.2^\circ$ (100 examples with step $1^\circ$). The testing data are generated at SNR$=-10$ and SNR$=0$ dB from $T=3,000$ and $T=1,000$ snapshots, respectively. The results of the DoA estimation without any knowledge of $K$ are depicted in Fig. \ref{fig:Exp6B}(a-c) for SNR$=-10$ dB and in Fig. \ref{fig:Exp6B}(d-f) for SNR$=0$ dB. In Fig. \ref{fig:Exp6B}(a), we observe that the DoAs are correctly estimated with only three FP (Type I error) occurrences. In Fig. \ref{fig:Exp6B}(b), we have two FP occurences, whereas in Fig. \ref{fig:Exp6B}(c), we have four FN (Type II error) occurrences. The confidence levels for each $K=1,2,3$ are $\bar{p}=0.88, 0.84$ and $0.71,$ respectively. Despite the low-SNR and the fact that the number of transmitting sources is unknown, the proposed CNN manages to estimate the (unknown) directions remarkably well. At higher SNRs (0 dB), in Figs. \ref{fig:Exp6B}(d-f), the results look even more promising. In particular, in Fig. \ref{fig:Exp6B}(d) and \ref{fig:Exp6B}(e) no errors occur in the number of estimated sources; the errors are due to the grid mismatch (quantization errors). In Fig. \ref{fig:Exp6B}(f), we observe only two FN errors. Of course, this second approach with the unknown number of sources can also be applied in the high-SNR regime (where we could get sufficient estimates with even smaller number of snapshots) and does not have to be limited to the low-SNR region only. However, we would not be able to overcome the grid mismatch error and thus, the results are not expected to improve much compared to those at 0 dB SNR.

\begin{table}
  \begin{center}
    \caption{CNN's DoA estimation results without knowledge on the number of transmitting sources.}
    \label{tab:Hausdorff_dis_vs_RMSE}
 \begin{tabular}{|c||c|c|c||c|c}
 \hline
 \multicolumn{5}{|c|}{SNR$=-10$ dB} \\
 \hline
    \textbf{Number of} & \textbf{Confidence} & \textbf{Mean} $d_{\text{H}}$ & \textbf{Max} $d_{\text{H}}$ & \textbf{RMSE}\\
    \textbf{sources $K$} & \textbf{level} $\bar{p}$ & $\mu ( d_{\text{H}})$ & $\max ( d_{\text{H}})$ & ($K$ known )\\
 \hline
	1 & 90\% & $0.20^\circ$ & $0.8^\circ$ & $0.20^\circ$\\
	2 & 74\% & $0.57^\circ$ & $5.8^\circ$ &  $0.41^\circ$\\
	3 & 71\% & $0.78^\circ$ & $10.6^\circ$ & $0.43^\circ$\\
 \hline
\end{tabular}\\
\begin{tabular}{|c||c|c|c||c|c}
 \hline
 \multicolumn{5}{|c|}{SNR$=0$ dB} \\
 \hline
    \textbf{Number of} & \textbf{Confidence} & \textbf{Mean} $d_{\text{H}}$ & \textbf{Max} $d_{\text{H}}$  & \textbf{RMSE}\\
    \textbf{sources $K$} & \textbf{level} $\bar{p}$ & $\mu ( d_{\text{H}})$ & $\max ( d_{\text{H}})$& ($K$ known )\\
 \hline
	1 & 90\% & $0.20^\circ$ & $0.2^\circ$ &  $0.20^\circ$\\
	2 & 77\% & $0.60^\circ$ & $5.8^\circ$ &  $0.43^\circ$\\
	3 & 70\% & $0.86^\circ$ & $10.6^\circ$ &  $0.40^\circ$\\
 \hline
\end{tabular}
  \end{center}
\end{table}

\section{Conclusions and Future Work}
\label{sec:conclusions}

In this paper, we introduced a deep convolutional neural network (CNN) with 2D filters for DoA prediciton in the low-SNR regime. In particular, we modeled the angle estimation as a multi-label classification task by considering an on-grid approach. The adoption of 2D convolutional layers enables the feature extraction from the multi-channel input data and the transfer of information to fully connected layers, leading to the robust DoA estimation in the low SNR. Two different training strategies are proposed: a) for a fixed and b) for a varying number of sources. The former is suitable in cases where the number of sources is known a priori, which is a typical assumption in the related literature. The proposed solution is compared against state-of-the-art methods and the Cram\'{e}r-Rao lower bound is also provided as benchmark. The performance of the proposed CNN is evaluated in terms of the DoA RMSE for off-grid angles under various setups: i) fixed and varying directions, ii) across a wide range of SNRs, iii) while varying the number of snapshots, iv) while varying the angular separation of the sources and v) in situations of SNR mismatches. The results indicate a) enhanced robustness, b) resilience to the estimation for a wide range of snapshots and c) ability to resolve closely spaced angles in the low SNR. 

Additionally, we introduced a training approach for a varying number of sources for application scenarios where the number of sources is unknown, which leads to a probabilistic method. The reported results indicate that the proposed CNN is able to successfully identify the (unknown) number of sources jointly with the DoAs with high probability. Moreover, the predicted angles are sufficiently good estimates (within the grid's resolution) to the true DoAs. Possible future research directions include deriving architectures with very high grid resolution (finer grid) for the identification of multiple targets.

%\section*{Acknowledgment}

\bibliographystyle{ieeetr}
\bibliography{IEEE_gp_bibliography}

%\begin{IEEEbiography}{Georgios K. Papageorgiou}
%\end{IEEEbiography}
%
%\begin{IEEEbiography}{Mathini Sellathurai}
%\end{IEEEbiography}

\end{document}